\documentclass{aa}
\usepackage[varg]{txfonts}

\usepackage{graphicx}
\usepackage{natbib}
\usepackage{amsmath}
\usepackage{multicol}
\usepackage{pdflscape}
\usepackage{threeparttable}
\usepackage[debugshow]{supertabular}
\makeatletter
\renewcommand*{\@fnsymbol}[1]{\ensuremath{\ifcase#1\or *\or \dagger\or \ddagger\or
     \mathsection\or \mathparagraph\or \|\or **\or \dagger\dagger
        \or \ddagger\ddagger \else\@ctrerr\fi}}
    \makeatother
\graphicspath{{figures/}}
\defcitealias{Saglia10}{S10}
\defcitealias{Opitsch18}{Paper I}
\begin{document}

\title{Stellar populations of the central region of M31.\thanks{This paper includes data taken at The McDonald Observatory of The University of Texas at Austin.} \thanks{This research was supported by the DFG cluster of excellence 'Origin and Structure of the Universe'.}\thanks{Tables \ref{tab:binid}, \ref{tab:Lick_indices} and \ref{tab:Stellar_populations} are only available in electronic form at the CDS via anonymous ftp to cdsarc.u-strasbg.fr (130.79.128.5) or via http://cdsweb.u-strasbg.fr/cgi-bin/qcat?J/A+A/} }

\author{R.P. Saglia \inst{1,2}
\and M. Opitsch \inst{1,2,3}
\and M.H. Fabricius \inst{1,2}
\and R. Bender \inst{1,2}
\and M. Bla\~na \inst{1}
\and O. Gerhard \inst{1}
}

\institute{Max Planck Institute for Extraterrestrial Physics, Giessenbachstr., D-85748, Garching, Germany
\and Universit\"ats-Sternwarte M\"unchen, Scheinerstr. 1, D-81679, Munich, Germany
\and Excellence Cluster Universe, Boltzmannstr. 2, D-85748, Garching, Germany}

\date{}

\abstract{{}{\textbf{Aims}: We continue the analysis of the data set of
    our spectroscopic observation campaign of M31, whose ultimate goal
    is to provide an understanding of the three-dimensional structure
    of the bulge, its formation history, and composition in terms of
    a classical bulge, boxy-peanut bulge, and bar contributions.} \\
  { \textbf{Methods}: We derive simple stellar population (SSP) properties,
    such as age metallicity and $\alpha-$element overabundance, from the
    measurement of Lick/IDS absorption line indices. We describe their
    two-dimensional maps taking into account the dust distribution in M31.} \\
  {\textbf{Results}: We found 80\% of the values of our age measurements are
    larger than 10 Gyr. The central 100 arcsec of M31 are dominated
    by the stars of the classical bulge of M31. These stars are old ($11-13$
    Gyr), metal-rich (as high as [Z/H]$\approx 0.35$ dex) at the
    center with a negative gradient outward and enhanced in
    $\alpha-$elements ([$\alpha/$Fe]$\approx 0.28\pm 0.01$ dex }). The
  bar stands out in the metallicity map, where an almost solar value
  of [Z/H] ($\approx 0.02\pm 0.01$ dex) with no gradient is observed
  along the bar position angle (55.7 deg) out to 600 arcsec from the
  center. In contrast, no signature of the bar is seen in the age and
  [$\alpha/$Fe] maps, which are approximately axisymmetric, delivering
  a mean age and overabundance for the bar and boxy-peanut bulge
  of 10-13 Gyr and 0.25-0.27 dex, respectively. The boxy-peanut bulge
  has almost solar metallicity ($-0.04\pm 0.01$ dex). The
  mass-to-light ratio of the three components is approximately
  constant at $M/L_V \approx 4.4-4.7 M_\odot/L_\odot$. The disk
  component at larger distances is made of a mixture of stars, as
  young as 3-4 Gyr, with
  solar metallicity and smaller $M/L_V$ ($\approx 3\pm0.1 M_\odot/L_\odot$).\\
  {\bf Conclusions:} We propose a two-phase formation scenario for the
  inner region of M31, where most of the stars of the classical bulge
  come into place together with a proto-disk, where a bar develops and
  quickly transforms it into a boxy-peanut bulge. Star
  formation continues in the bulge region, producing stars younger than
  10 Gyr, in particular along the bar, thereby enhancing its metallicity. The
  disk component appears to build up on longer timescales.}


\maketitle

\section{Introduction}

Because of its proximity, M31 is the best case after the Milky Way
(MW) to study the detailed evolutionary history of a large spiral
galaxy.  There is strong evidence from photometry for the presence of
a barred bulge in M31 \citep{Lindblad56, Beaton07} and N-body models;
see, for example, \citet{Athanassoula06} and \citet[hereafter
B17]{Blana16a}.  In \citet[hereafter Paper I]{Opitsch18}, we present
arguments for the presence of a bar based on the kinematics of the
stars, ionized gas, and gas fluxes, backing the M31 model of
B17. \citet{Blana18} have refined this model using the NMAGIC
technique of \citet{deLorenzi07}, adapted to barred disk galaxies as
in \citet{Portail17}, to fit the kinematic maps presented in Paper I.
In their scenario, M31 belongs to the category of composite bulges
\citep{Erwin15}, harboring a classical bulge (hereafter CB) component
dominating the dynamics inside 100 arcsec from the center and
containing one-third of the total bulge mass (B17), and a boxy-peanut
(hereafter B/P) generated by the buckling of a bar presently detected
at a position angle of $PA_{bar}=55.7$\degr and with a projected
length of $\approx 600$\arcsec. Outside this region the disk component
of M31 starts to dominate.  On the other end, the photometric
$R^{1/4}$ plus exponential decomposition along the major axis of
\citet{Kormendy99} could suggest that a (possibly young and) rotating
disk component is also present inside the 600\arcsec region.

Furthermore, only determining the ages and metallicities of the
stellar populations of the bar, the classical and B/P bulge can
clarify when these different structures formed and constrain their
mass-to-light ratios. Presently, the modeling approach of
\citet{Blana18} assumes that they all have the same uniform value and
this assumption needs verification. Finally, a plausible formation
scenario should emerge from a joint dynamical and stellar population
modeling approach. In the long run, the origin of the microlensing
events observed toward the bulge of M31 \citep{Lee12} might be
understood. Some of these could originate from the dark matter halo of
M31, if a fraction of it is made of compact massive objects. To reach
this conclusion we need a well constrained three-dimensional model of
the stellar populations of the bulge region of M31 that allows us to
estimate the number of microlensing events toward M31 due to
self-lensing \citep{Riffeser08}.

Bars are thought to be the major drivers of the so-called secular
evolution, which is the slow rearrangement of energy and mass in
galaxies, as opposed to the violent and rapid processes of
hierarchical clustering and merging (see, e.g., \citealp{Kormendy04,
  Athanassoula13, Kormendy13, Sellwood14}).  Through their ability to
transport angular momentum and gas, bars influence the inner regions
of galaxies \citep{Knapen95, Bureau99, Sakamoto99,
  Fathi03, Chung04, Cheung13}.  They also affect the outer regions by
redistributing the stellar component \citep{Gadotti01,Roskar12}, by increasing
radial motions and reshuffling the stellar content, which leads to
flatter radial gradients in the stellar populations \citep{Minchev10,
  DiMatteo13, Kubryk13}.

So far there is no conclusion yet on the impact of the presence of a
bar on metallicity gradients.  Several studies have found that the
stellar metallicity seems to remain unchanged in the central parts in
the presence of a bar \citep{Friedli94, Coelho11, Williams12a,
  Cacho14, Cheung15}, while others found a mild increase
\citep{Moorthy06, Perez11}. \citet{Seidel16} measured larger gradients
in metallicity for barred galaxies than in unbarred galaxies with a
steeper gradient within 0.06 bar lengths. 
  \citet{DiMatteo2015},\citet{DiMatteo2016},\citet{Athanassoula17} and 
\citet{Fragkoudi17}
  emphasize the role of the thick disk to explain the metallicity
  distribution of the MW. \citet{Debattista17} and
  \citet{Gonzalez17} argued that the X-shaped structure of
  B/P-shaped bulges are preferentially populated by
  metal-rich stars because of  kinematic fractionation. \\

Regarding the influence of bars on stellar populations in the outer
regions of galaxies, there is also no consensus in the literature.
Some studies find no systematic difference between barred and unbarred
galaxies \citep{Perez07,Perez09}, while others find flatter age and
metallicity gradients along the bar than compared to those along the
disk \citep{Sanchez-Blazquez11, Seidel16}.

In this paper we exploit our VIRUS-W spectroscopic observations of M31
(see Paper I) to study the stellar populations of the components of
its central regions (the bar, classical and B/P bulge, and disk)
and address the open questions discussed above.  \citet[hereafter
S10]{Saglia10} measured the stellar populations of the central region
of M31 along six position angles deriving old ages, solar metallicity,
and slight $\alpha-$elements overabundance. We complete and
revise the picture by presenting two-dimensional maps of these
quantities. 

Our observations are complementary to the study of
 \citet{Dong2018}, who  analyzed the color-magnitude diagrams of the inner
  $\approx 200$ arcsec bulge region of M31. These authors detected a widespread
  population of stars younger than 5 Gyr, but  80\% of these stars have stellar mass outside 100 arcsec from the center that is older than
  8 Gyr and metal rich ([Fe/H]$\sim0.3$).

We observed M31 for 14 nights with the integral field
spectrograph \texttt{VIRUS-W} attached to the 2.7 m telescope at the
McDonald Observatory, Texas, covering the bulge area with a filling
factor of one-third and sampling the disk along six different directions,
reaching approximately one scalelength along the major axis.
The integral field unit \texttt{VIRUS-W} of consists of 267 fibers that are
arranged in a rectangular pattern. The wavelengths covered range from
4800 \AA\ to 5400 \AA,  spectral resolution is $R=9000$, which
corresponds to $\sigma_{inst}$=15$km s^{-1}$. A detailed description
of the observations can be found in Paper I.

We rebinned the 56.000 spectra to 7562 binned spectra to
increase the signal-to-noise ratio in the outer parts and reach
  the minimum value of 30. These binned spectra were fitted with
\texttt{pPXF} \citep{Cappellari03} and \texttt{GANDALF}
\citep{Sarzi06} to obtain the kinematics of the stars, ionized gas, gas fluxes, and ionization
mechanisms. This is described in detail in \citetalias{Opitsch18} 
  and results in 6473 velocity determinations.  We focus on the
measurement of the absorption line indices and the stellar
populations for these 6473 spectra.

The paper is structured as follows. In Sect.
\ref{sec:StellarPopulations}, we describe the measurement of the
absorption line indices and the derivation of the stellar population
properties. Sect. \ref{sec:Discussion} discusses the stellar
population properties of the classical and B/P bulge components,
bar, and disk, proposing a possible formation scenario. We
summarize our findings in Sect. \ref{sec:Conclusions}.  The appendix
shows the comparison with our previous measurements of \citetalias{Saglia10}
and gives examples of the format of the tables presenting the data and
available online. As in Paper I, we adopt a distance to M31 of 
$0.78\pm 0.04$ Mpc. At this distance 1 arcsec corresponds to 3.78 pc.

\section{Methods}
\label{sec:StellarPopulations}

In this section, we present the measurement of absorption line
strengths in the Lick/IDS system \citep{Burstein84, Faber85,
  Burstein86, Worthey94} and the comparison to simple stellar population (SSP) models \citep{Maraston98, Maraston05, Thomas03, Thomas11} 
to obtain the age, metallicity and
[$\alpha$/Fe] overabundance of the stellar populations.  As described in
\citetalias{Opitsch18}, we used \texttt{pPXF} \citep{Cappellari03} to
fit the stellar kinematics of the stars and \texttt{GANDALF}
\citep{Sarzi06} to fit the kinematics and line fluxes of the ionized gas,
mainly using the [OIII]$\lambda$5007 line.

In order to interpret the features visible in the maps of the Lick
indices presented below, we took the distribution
of the dust in M31 into account. \citet{Blana18} have produced a map projected onto the
plane of the sky of the ratio $R_{dust}$ of the three-dimensional
model luminosities measured with and without dust using the dust maps
of \citet{Draine2014}, converted to $A_V$ following
\citet{Draine2007}. According to this map, the light integrated along
the near side of M31 is dominated by the outer part of the disk, since
the rest is highly absorbed ($R_{dust}<0.8$).  In the far side of M31,
where $R_{dust}>0.8$ the light of the bulge region and the inner part
of the disk dominates. Fig. 1 of B17 and \citet{Blana18} illustrate
these geometric situations; see also \citet{Dalcanton15}.

\begin{figure}
\resizebox{\hsize}{!}{\includegraphics{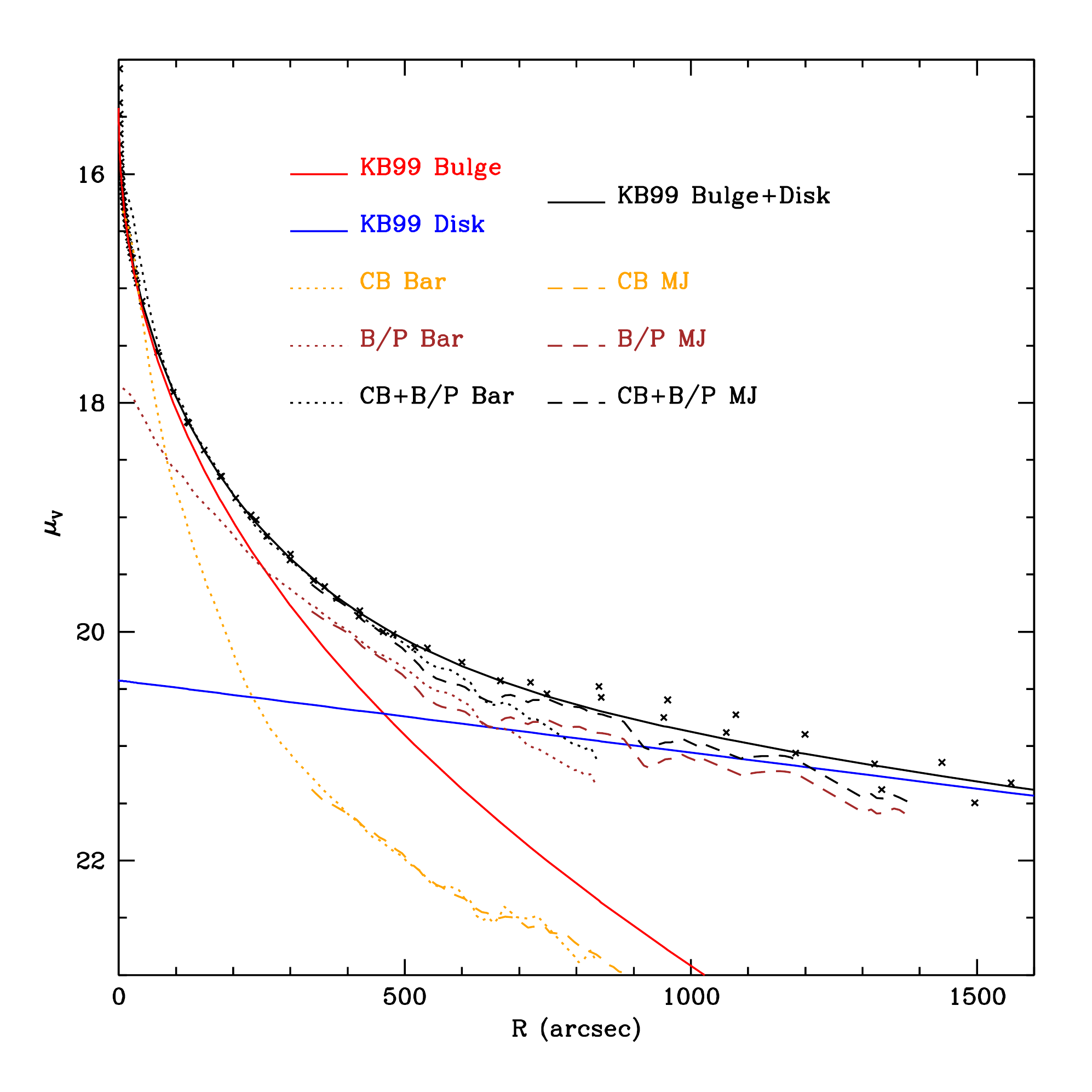}}
\caption[Kormendy profile]{ V-band major axis profile (crosses),
  bulge (red line), and disk (blue) decomposition of
  \citet{Kormendy99}. The black line shows the sum of the two
  components.  The orange and brown dotted lines show the CB and B/P
  profiles, respectively, in the bar region (see
  Fig. \ref{fig:regions}) of \citet{Blana18}.  The
  black dotted line shows the sum of the two components. The orange
  and brown dashed lines show the CB and B/P bulge and disk profiles along the
  position angle of the major axis; the black dashed line indicates their sum. }
\label{fig:Kormendy}
\end{figure}

As a second ingredient, we need to define quantitatively the different
structures present in M31. We distinguish between a disk and a bulge
photometric region using the decomposition of \citet{Kormendy99}
adopted in Paper I and shown in Fig. \ref{fig:Kormendy}. Accordingly,
the photometric bulge region is the region of M31 where the local
bulge-to-total ratio of M31 is higher than 0.5.  Following B17 we call
B/P bulge the $1200''\times540''$ rectangular portion of the sky
around the center of M31 with position angle $PA_{bar}=55.7$ degrees,
excluding the central $100''\times100''$; this area is dominated by the
light of the CB. We call bar major axis the
$1200''\times100''$ rectangular portion with the same $PA_{bar}$, and
bar minor axis the $270''\times100''$ rectangular portion with
$PA=PA_{bar}+90$.  Figs. \ref{fig:schema} and \ref{fig:regions}
provide a graphical representation of our definitions.

\subsection{Lick indices}

In Paper I we describe how we removed the fitted emission lines from
the binned spectra before our derivation of line-of-sight velocity
distributions. Using these cleaned, flux calibrated spectra we now
measure the Lick indices.  For the Lick/IDS system we use the band
definitions of \citet{Trager98}.  The indices that fall into our
spectral range are H$\beta$, Mg b and the iron indices Fe5015, Fe5270,
Fe5335 and Fe5406. The indices H$\beta$, Mg b and Fe5015 are
  affected by emission; on average, the corrections amount to
  $0.3\pm0.07$ \AA, $-0.1\pm 0.07$ \AA\ and $0.5\pm 0.07$ \AA,\,
  respectively. These values are very similar to those reported by
  \citet{Saglia10}. No correlations are seen between the index values and
  the size of the corrections. 

Errors on the index measurements are estimated using a Monte Carlo
approach. For each spectrum, 1000 representations with
added random Gaussian noise are created and the width of the Gaussian is 
the noise value of the binned spectra from \citetalias{Opitsch18}. 
The typical errors are around 0.1-0.2 \AA .
\\

We make sure that we are on the Lick/IDS system by comparing our
measured values to those by \citetalias{Saglia10}, which have been
calibrated to the Lick/IDS system.  The comparison plots are shown in
Sect. \ref{sec:Saglia_indices} in the Appendix and both datasets agree
very well. In particular, the agreement at the level of 0.1 \AA\
  between the H$\beta$, Fe5015, and Mgb indices is reassuring of the
  consistency of the removal of the emission lines \citep[see Fig. 8 of][]{Saglia10}.

The resulting maps for the Lick indices are plotted in
Figs. \ref{fig:LickHbeta_map} to \ref{fig:LickFe5406_map}; they have
been smoothed with a two-dimensional Gaussian filter with
$\sigma=10$\arcsec \ to reduce the noise in the maps and make
the trends clearer. Their tabulated values are available online, a
small part of which are presented in Table \ref{tab:Lick_indices}
in the Appendix to clarify the format.

\begin{figure}
\resizebox{\hsize}{!}{\includegraphics{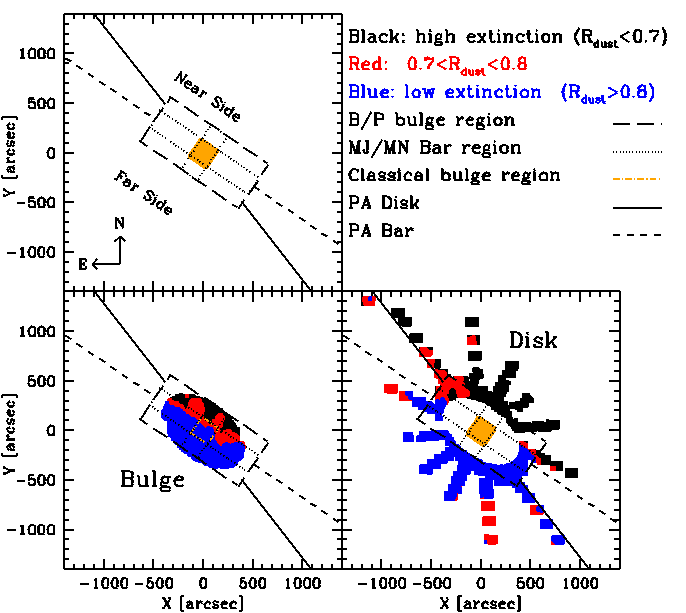}}
\caption[Definition of regions] {Schematic definition of the
  various regions of M31 discussed in
  Sect. \ref{sec:StellarPopulations}. Regions of low extinction
  ($R_{dust}>.8$) indicated in blue, regions of high extinction
  ($R_{dust}<.7$) in black, and the remaining in red.  The CB region is indicated in orange. North is to the top, east to the
  left.}
\label{fig:schema}
\end{figure}

\begin{figure}
\resizebox{\hsize}{!}{\includegraphics{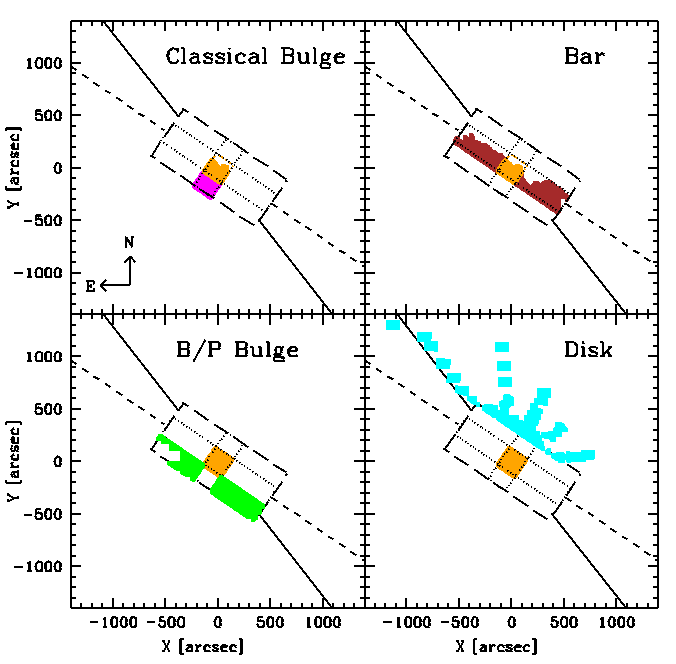}}
\caption[Definition of regions] {Schematic definition of the
  various regions of M31 discussed in
  Sect. \ref{sec:StellarPopulations}. The different colors show the
  points used in the derivation of Figs. \ref{fig:GradientMN} to
  \ref{fig:GradientDisk}. The inner (100 arcsec) region of the
  CB is indicated in orange, the outer region
  perpendicular to the bar in magenta. The bar region outside the
    CB is represented in brown, the B/P region in green, and
    the disk in cyan. North is to the top, east to the left.}
\label{fig:regions}
\end{figure}

\begin{figure}
\resizebox{\hsize}{!}{\includegraphics{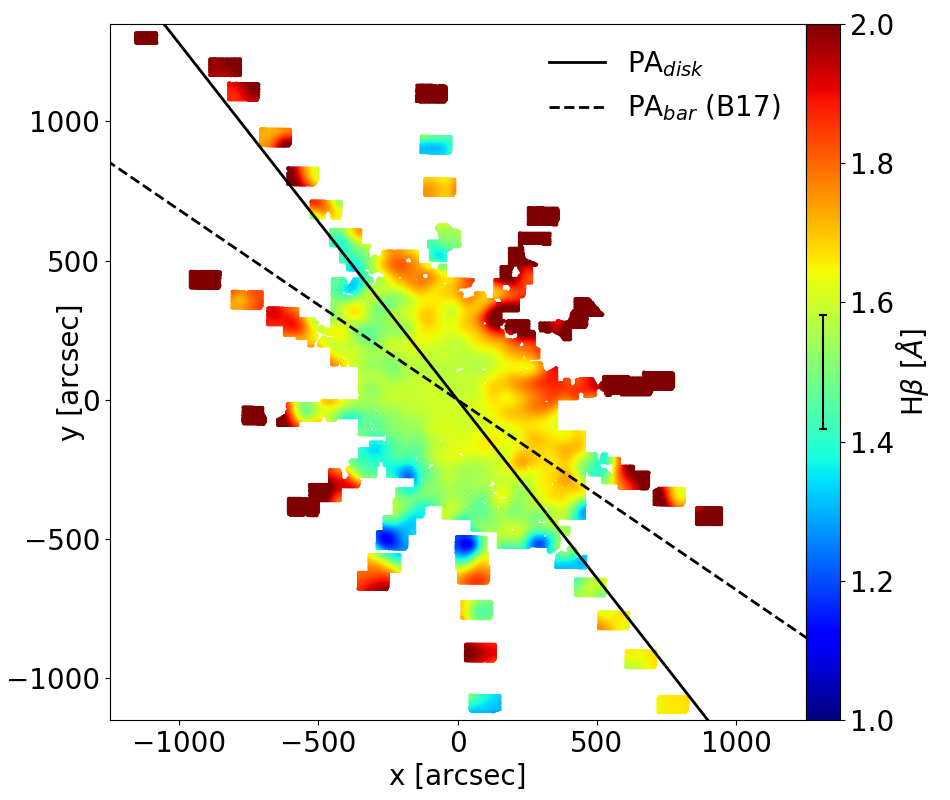}}
\caption[Map of Lick index H$\beta$]{Smoothed map of the Lick index
  H$\beta$ with the position of the disk major axis (solid line) and
  the bar major axis (dashed line).  The median error bar of the
  individual measurements is plotted in the color bar. At the assumed
  M31 distance of 0.78 kpc, 1 arcsec corresponds to 3.78 pc.}
\label{fig:LickHbeta_map}
\end{figure}

\begin{figure}
\resizebox{\hsize}{!}{\includegraphics{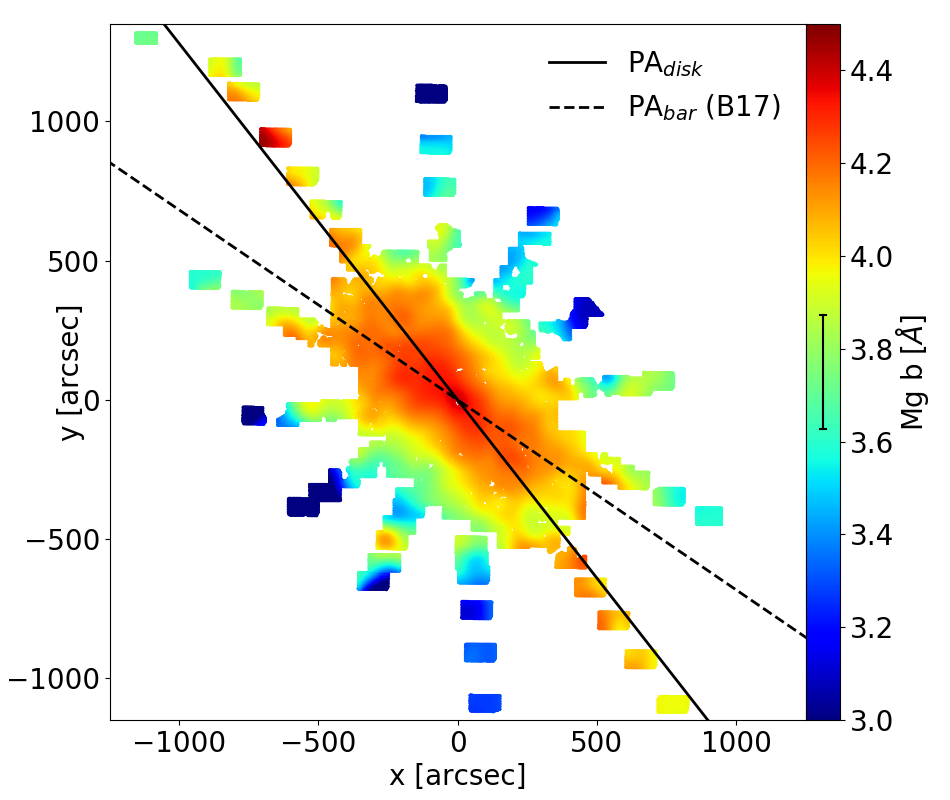}}
\caption[Map of Lick index Mg b]{Smoothed map of the Lick index Mg b. The lines are
 as in Fig. \ref{fig:LickHbeta_map}. }
\label{fig:LickMgb_map}
\end{figure}
\begin{figure}
\resizebox{\hsize}{!}{\includegraphics{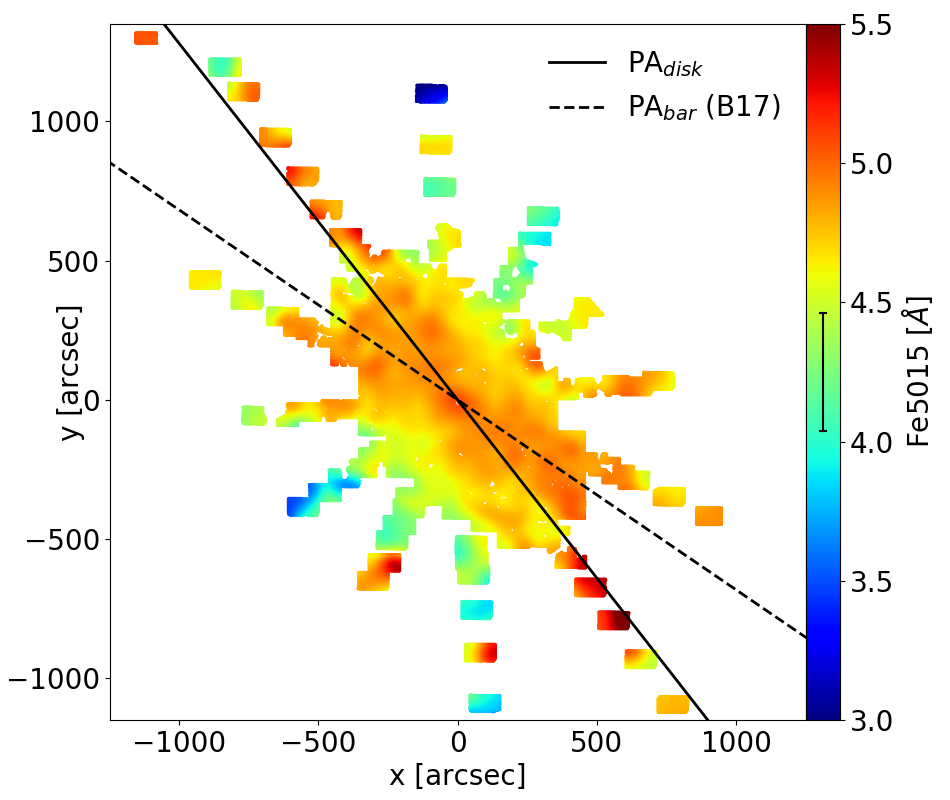}}
\caption[Map of Lick index Fe5015]{Smoothed map of the Lick index Fe5015. The lines are as in Fig. \ref{fig:LickHbeta_map}.}
\label{fig:LickFe5015_map}
\end{figure}
\begin{figure}
\resizebox{\hsize}{!}{\includegraphics{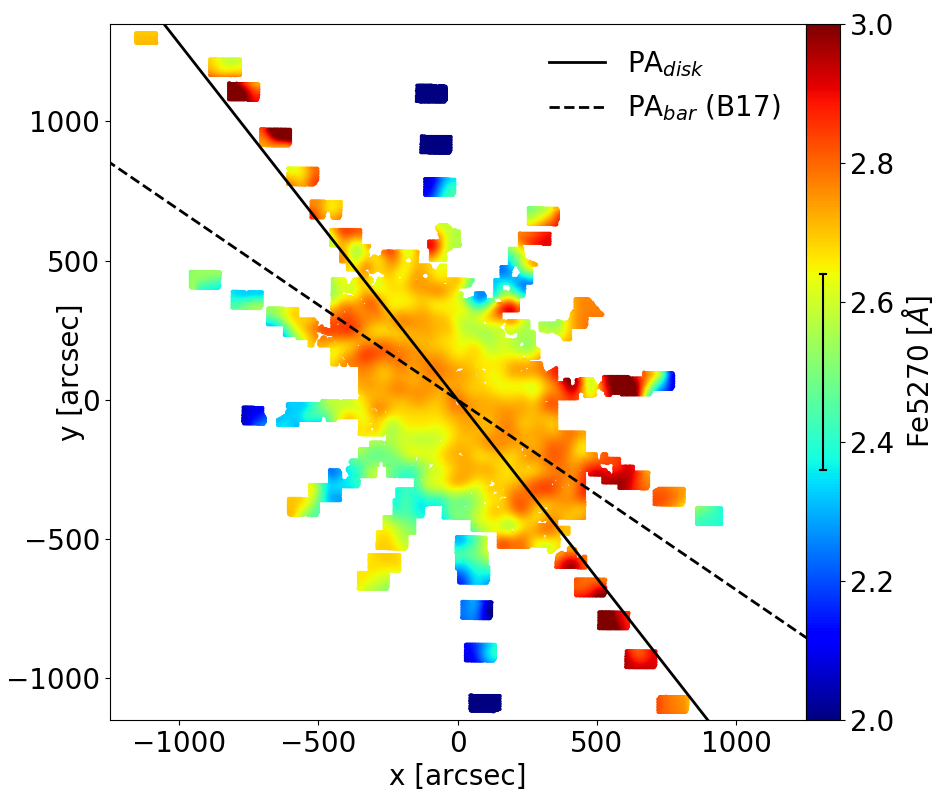}}
\caption[Map of Lick index Fe5270]{Smoothed map of the Lick index Fe5270. The lines are as in Fig. \ref{fig:LickHbeta_map}.}
\label{fig:LickFe5270_map}
\end{figure}
\begin{figure}
\resizebox{\hsize}{!}{\includegraphics{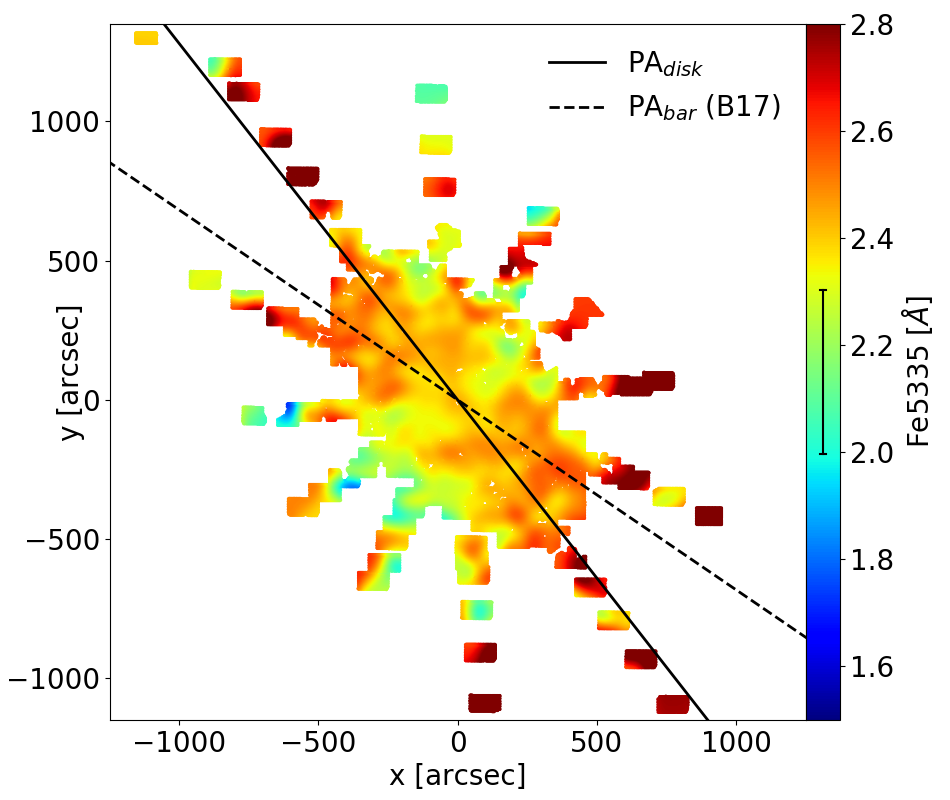}}
\caption[Map of Lick index Fe5335]{Smoothed map of the Lick index Fe5335. The lines are as in Fig. \ref{fig:LickHbeta_map}.}
\label{fig:LickFe5335_map}
\end{figure}
\begin{figure}
\resizebox{\hsize}{!}{\includegraphics{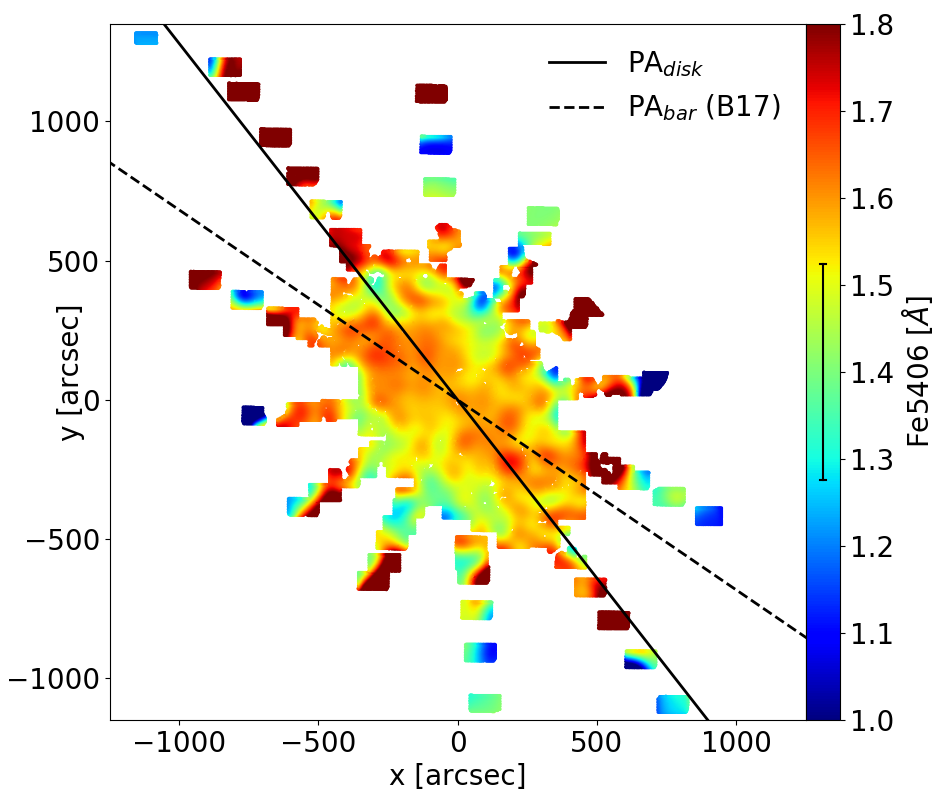}}
\caption[Map of Lick index Fe5406]{Smoothed map of the Lick index Fe5406. The lines are as in Fig. \ref{fig:LickHbeta_map}.}
\label{fig:LickFe5406_map}
\end{figure}

The overall distribution of the H$\beta$ values is asymmetric; see
Fig. \ref{fig:LickHbeta_map}, where higher values are on the 
  northwestern side and lower values on the southeastern side
of the galaxy. A comparison with Fig. \ref{fig:schema} shows that in
the first case we are probing that near side of M31, where the dusty
disk largely screens the inner high density regions of the bulge and
where we expect to find the younger population of stars of the
external regions of the disk, with higher values of H$\beta$. In the
second case we are looking through the relatively unextincted
classical, B/P, and bar regions, where an older population of stars
with lower values of H$\beta$ could be expected.

The Mg b index, as an $\alpha$-element important for the measurement
of the [$\alpha/$Fe] overabundance, is plotted in
Fig. \ref{fig:LickMgb_map}. The overall appearance of the map shows a
clear enhancement in the B/P bulge region, which has particularly high
values in the CB region and a strong negative gradient
perpendicular to the bar and a milder gradient parallel to the bar (see
Fig. \ref{fig:regions}). In the disk region the Mg b index is smaller.

The iron indices show a similar pattern to the Mg b index map, which is
also elongated and enhanced along the bar direction. There is a hint
of a central pinching orthogonal to the bar, reminiscent  of
  what is discussed in \citet{Debattista17}, \citet{Fragkoudi17}, and
  \citet{Valpuesta13}. A slight pinching is also observed in the MW bulge \citep{Gonzalez13}.

\subsection{Stellar population analysis}
\label{spanalysis}

We adopted the method by \citetalias{Saglia10} to measure stellar
population properties.  We assume that a spectrum of M31 is well
represented by one single stellar population, which is chemically
homogeneous and in which all stars have the same age.  As
  discussed above, several components are present in the bulge region
  of M31. Interpreting mixed stellar populations with SSPs biases the
  inferred stellar population properties toward the youngest
  component present. In
  Sect. \ref{sec:Discussion} we discuss the size of this effect. \\
For these populations we take  SSP models
from \citet{Maraston98, Maraston05}, which are combined with a
\citet{Kroupa01} initial mass function \ (IMF) and Lick indices models that take
into account the [$\alpha$/Fe] overabundance by \citet{Thomas03,
  Thomas11}. This is a reasonable assumption, given that
\citet{Zieleniewski15} found that a Chabrier IMF is appropriate for the
bulge of M31. We note that only the M/L ratios and not the line indices
considered in this work are sensitive to this choice.  \\

Since the original models by \citet{Maraston98, Maraston05} and
\citet{Thomas03,Thomas11} are computed for a relatively coarse grid of
age, [Z/H] and [$\alpha/$Fe], we interpolated these onto a finer grid,
ranging in age from 0.1 to 15.0 Gyr with a step size of 0.1 Gyr, in
metallicity from -2.25 to 0.67 with a step size of 0.02, and in
[$\alpha$/Fe] overabundance from -0.3 to 0.5 with a step size of 0.05.\\

We also considered the models of \citet{Vazdekis2015}, which are
  provided for two values of the [$\alpha$/Fe] overabundance (0 and
  0.4 dex). The resulting stellar population parameters are similar to
  what is discussed below, in particular the inability of reproducing a
  large percentage of H$\beta$ values without hitting the maximum
  allowed age (in this case 14 Gyr). Hereafter we focus on the models
  of \citet{Thomas11}, given that they cover a larger range of
  [$\alpha$/Fe] overabundances.

The measured Lick index value for a binned spectrum $n$ is compared to
those in the model grid. For each gridpoint $i$, $\chi^2(i)$ is
calculated
\begin{align}
\label{eq:chi}
 \chi^2_n(i)&=\Delta {\rm H}\beta_n(i)^2+\Delta {\rm Mg} b_n(i)^2 + +\Delta {\rm Fe5015}_n(i)^2+\Delta {\rm Fe5270}_n(i)^2 \\
           & +\Delta {\rm Fe5335}_n(i)^2 +\Delta {\rm Fe5406}_n(i)^2. \nonumber
\end{align}
$\Delta Index_n(i)^2$ is
\begin{equation}
\Delta Index_n(i)^2 =\left(\frac{Index_{measured, n}- Index_{grid}(i)}{dIndex_{measured, n}} \right)^2.
\end{equation}

We computed solutions for both the \citet{Thomas03} and
\citet{Thomas11} models.  We find that our measured values for Fe5015
are reproduced well by the models of \citet{Thomas11}, but not by the
models of \citet{Thomas03}, which on average give Fe5015 values larger by 0.5 \AA\ than the measured data.  We therefore adopted the \citet{Thomas11}
models for the following analysis and derive solutions based on the
\citet{Thomas03} without considering the Fe5015 index to
compare with the results of S10 presented in the
Appendix. The grid index $i$ for which $\chi^2(i)$ is minimal is
determined and the values for  age$_{fit}(i)$, metallicity
[Z/H]$_{fit}(i),$ and overabundance [$\alpha/$Fe]$_{fit}(i)$ are then
taken as the values for the stellar populations. Since we do not
extrapolate the SSP model grids,  often (see below) the minimum
value is right at the edge of the grid. Using the age and metallicity
derived above we computed the corresponding $M/L_V$ ratios by
  interpolating the Kroupa IMF table of \citet{Maraston98,
    Maraston05}.  The values (one per binned spectrum) of the measured
  age, [Z/H], [$\alpha/$Fe], and $M/L_V$ are provided in tabular form
  online; an example that explains the format is presented in Table
  \ref{tab:Stellar_populations} in the Appendix.
  Fig. \ref{fig:distribution} shows their histograms and cumulative
  distributions. The formal statistical errors are estimated by
quoting the model range of values for which
$\chi^2_n - min(\chi^2_n) \leq 1$, with a minimum error on the age of
0.1 Gyr, on [Z/H] and [$\alpha/$Fe] of 0.01 dex, and on $M/L_V$ of
0.1 $M_\odot/L_\odot$. The average statistical errors on age, [Z/H],
  [$\alpha/$Fe], and $M/L_V$ are 1 Gyr, 0.04 dex, 0.02 dex, and 0.3
  $M_\odot/L_\odot$, respectively.\\

\begin{figure}
\resizebox{\hsize}{!}{\includegraphics{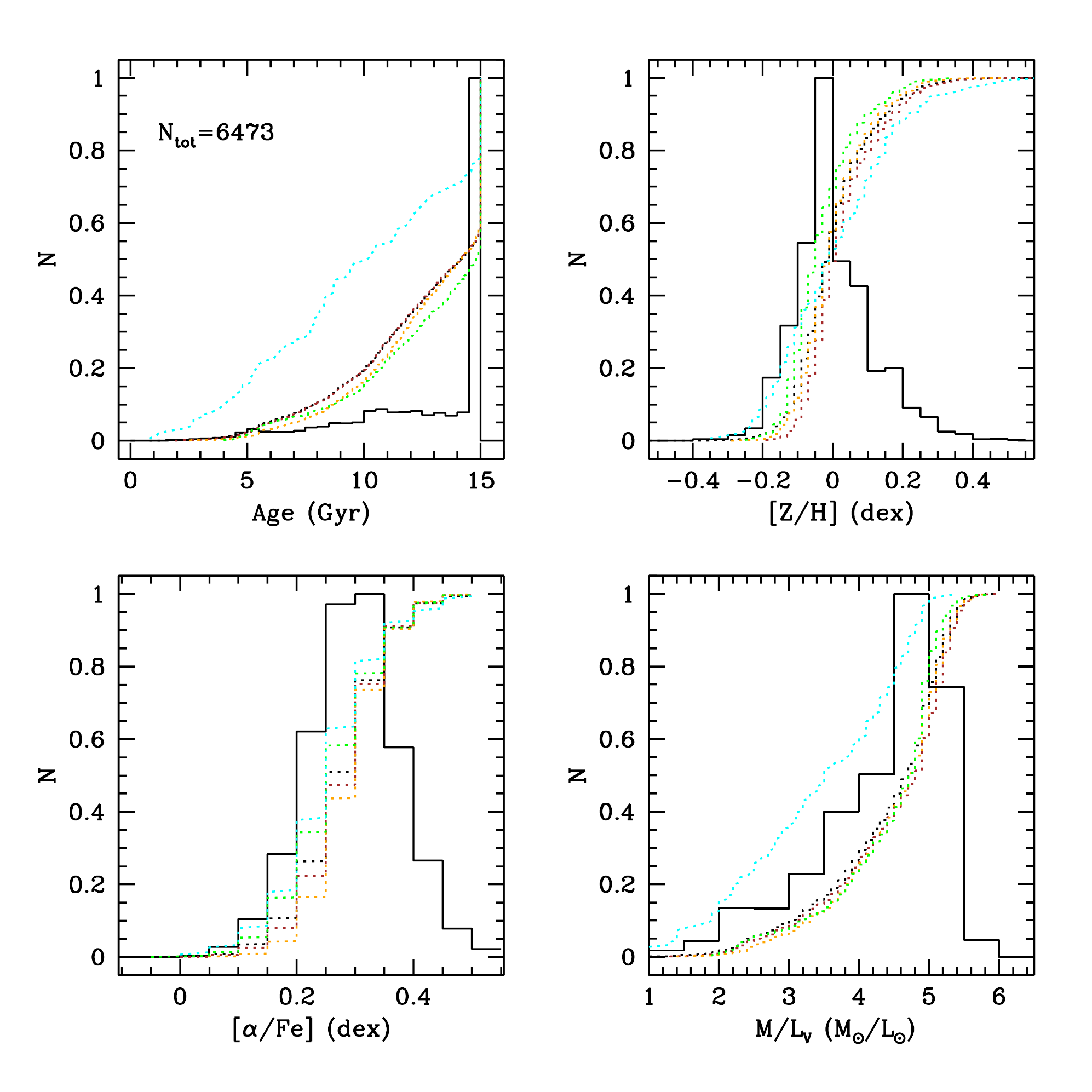}}
\caption[Distribution]{Histograms (full lines) and cumulative
  (dotted lines) distributions of the measured values of the age,
  [Z/H], [$\alpha/$Fe], and $M/L_V$. The orange, brown, green, and
  cyan dotted lines show the cumulative distributions of the CB bulge,
  bar, B/P bulge, and disk regions, respectively (see
  Fig. \ref{fig:regions}). }
\label{fig:distribution}
\end{figure}

The agreement between the measured index values and the model values
of \citet{Thomas11} is good; the median difference {\it
  Data-Model} is 0.11 \AA\ for Fe5335 (worst case), and -0.01 \AA\
for H$\beta$ (best case).  However, H$\beta$ model values as low
  as or lower than 1.6 \AA\ are obtained only for old, metal, and
  iron-rich SSPs, while our measurements require old, metal-rich, and
  $\alpha-$element overabundant SSPs.  Approximately half of our
  measured values are smaller than  1.6 \AA\ by on average 0.2 \AA. As
  discussed above, it is unlikely that this is because of a systematic
  oversubtraction of emission components.  This means that for almost
  40\% of the measurements (see Fig. \ref{fig:distribution}) the
  best-fitting age is the maximum allowed by the grid, i.e., 15 Gyr (larger
  than the age of the Universe of 13.8 Gyr), where our formal statistical
  errors are clearly underestimated. The remaining 60\% are
  distributed almost uniformly between 5 and 15 Gyr, of which 2.5\% of
  the measurements give ages smaller than 5 Gyr.  As a consequence,
  we overestimated the true ages of the stellar populations of M31;
  capping the measured ages to 13.8 Gyr, we reduced the overall average
  age by 0.6 Gyr from 12.6 to 12 Gyr. If instead we scale down every
  measured age by 13.8/15, we reduce the overall average age by 0.9
  Gyr.

  The impact on the derived metallicities and overabundances is
  however very limited. The distributions of measured metallicities
  and overabundances (see Fig. \ref{fig:distribution}) are
  approximately Gaussians. Allowing a maximum age of the models of
  13.8 increases the best-fitting metallicities of the affected
  spectra on average by only 0.026 dex (the overall average increases
  by 0.01 dex), well within the quoted errors, while the
  overabundances are essentially unchanged. We conclude that while the
  metallicites and overabundances are robust, we can trust the age
  rankings more than their absolute values.

  The distributions of the $M/L_V$ values is more skewed; a low value tail corresponds to the flat age distribution
  part. Under the effect of the age capping the derived mass-to-light
  ratios in the V band are reduced on average by
  $\approx -0.3 M_\odot/L_\odot$, approximately the size of the
  errors.

\begin{figure}
\resizebox{\hsize}{!}{\includegraphics{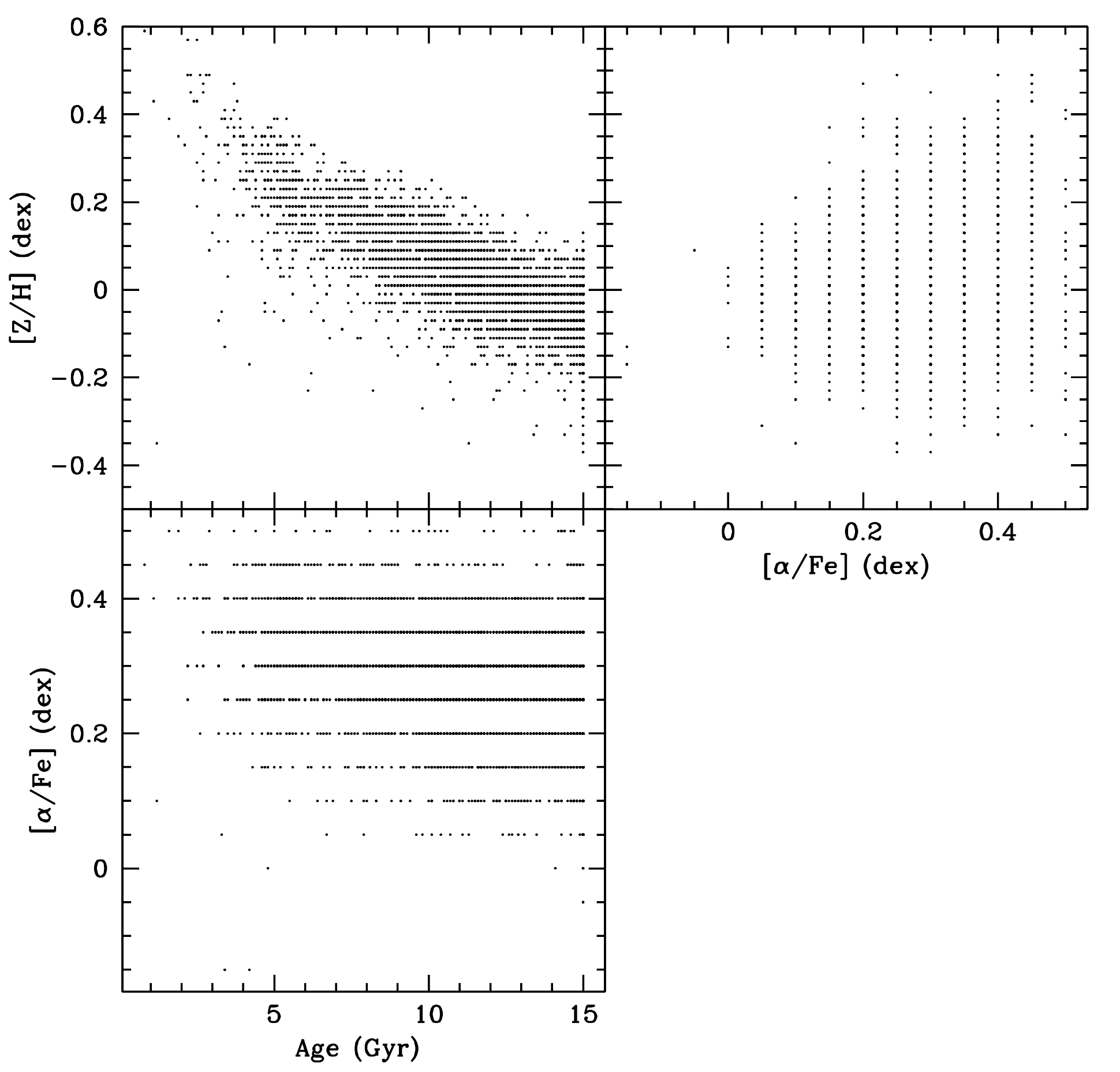}}
\caption[Correlation]{Correlations between the measured values of the 
age, [Z/H],  [$\alpha/$Fe], and $M/L_V$.}
\label{fig:correlation}
\end{figure}

When computed with the \citet{Thomas03} models, ages are 0.7 Gyr
smaller, metallicities are 0.04 dex larger, and [$\alpha/$Fe]
overabundances are 0.05 dex smaller than the median values derived
using the \citet{Thomas11} models.
Fig. \ref{fig:correlation} shows the correlations between age,
  [Z/H], and [$\alpha/$Fe].  Metallicities are broadly anticorrelated
  with ages and on average solar; $\alpha-$elements are almost always
  overabundant (on average by 0.27 dex) with respect to the solar
  values.

\begin{figure}
\resizebox{\hsize}{!}{\includegraphics{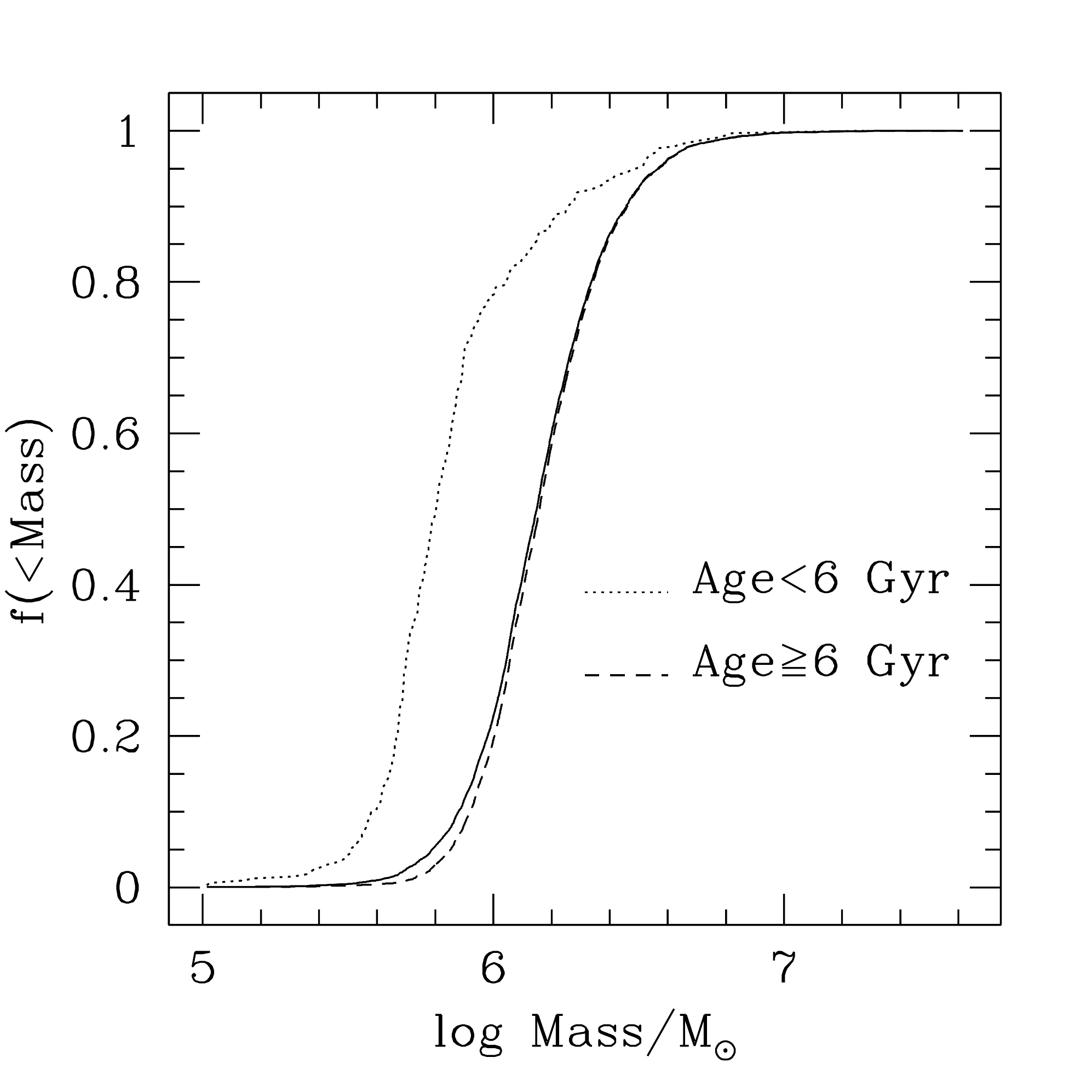}}
\caption[Correlation]{Cumulative distribution of the stellar
  masses sampled by our spectra (full line). The dotted line shows the
  cumulative distribution of the stellar masses of the spectra dated
  less than 6 Gyr, and the dashed line represents those dated more than 6 Gyr.}
\label{fig:Masscum}
\end{figure}

Using the $M/L_V$ derived above (see Fig. \ref{fig:distribution})
  we can translate the light sampled by each spectrum and estimated
  from the photometry of \citet{Kormendy99} into the corresponding
  stellar mass. Fig. \ref{fig:Masscum} shows that every spectrum
  samples at least $10^5M_\odot$ and that 80\% of our spectra sample
  at least $10^6M_\odot$. For the 5\% of spectra dated less than 6
  Gyr, 78\% sample less than $10^6 M_\odot$. These percentages of
  spectra sampling less than $10^6 M_\odot$ are probably
  overestimates, since the presence of young populations (see
  discussion below) can reduce the measured $M/L_V$ by up
  to 70\% with respect to the true value.  \citet{Bruzual2017} showed
  that a stellar mass of $10^6M_\odot$ suffices to average out the
  noise due to the stocastic sampling of the IMF and that below
  $10^5M_\odot$ the noise becomes prohibitively high.

\begin{figure}
\resizebox{\hsize}{!}{\includegraphics{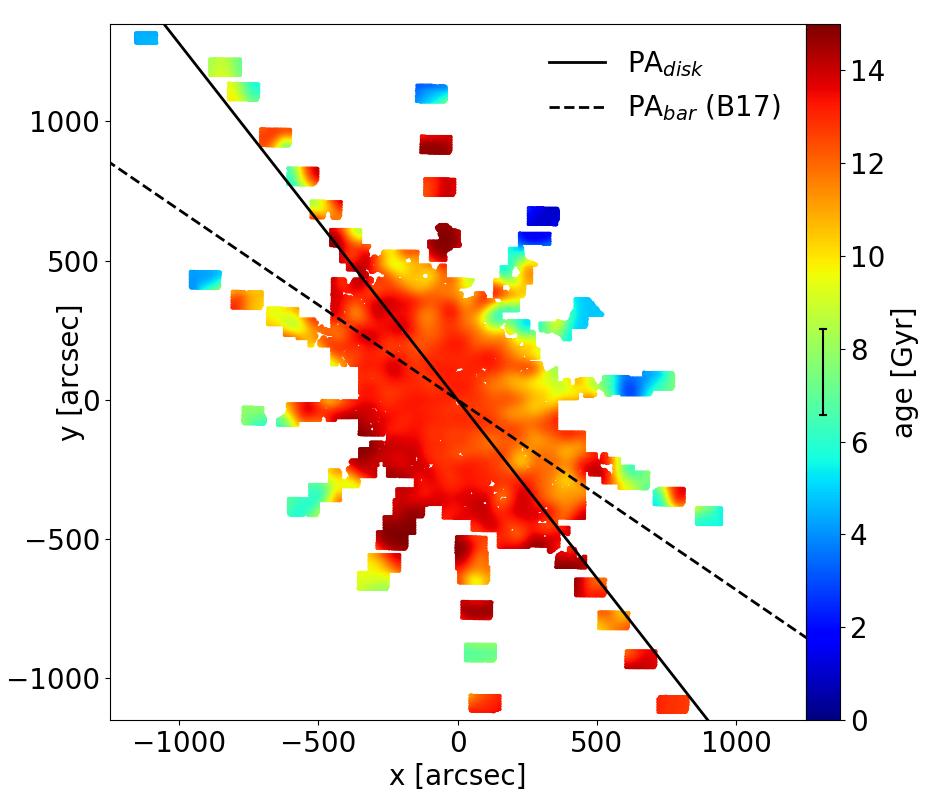}}
\caption[Age map]{Mean age map. The lines are as in Fig. \ref{fig:LickHbeta_map}. See Figs. \ref{fig:schema} and \ref{fig:regions} for a schematic description of the geometry of the galaxy, its components, and the distribution of the dust. }
\label{fig:age}
\end{figure}

\begin{figure}
\resizebox{\hsize}{!}{\includegraphics{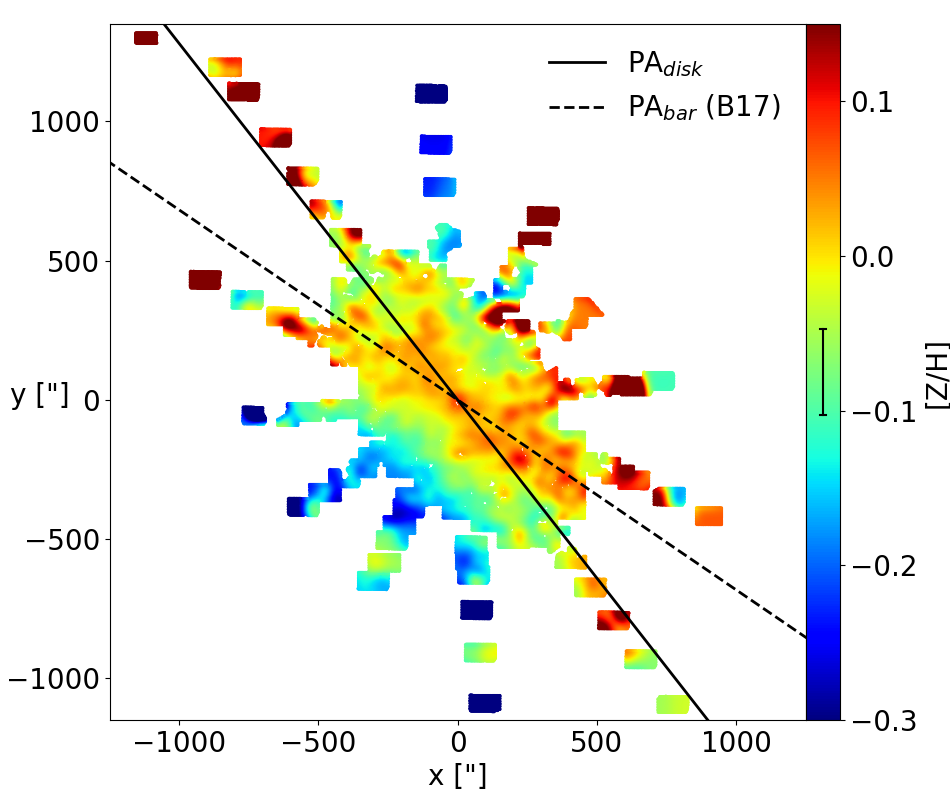}}
\caption[Metallicity map]{Mean metallicity map. The lines are as in 
Fig. \ref{fig:LickMgb_map}. See Figs. \ref{fig:schema} and \ref{fig:regions} for a schematic description of the geometry of the galaxy, its components, and the distribution of the dust.}
\label{fig:ZH}
\end{figure}

\begin{figure}
\resizebox{\hsize}{!}{\includegraphics{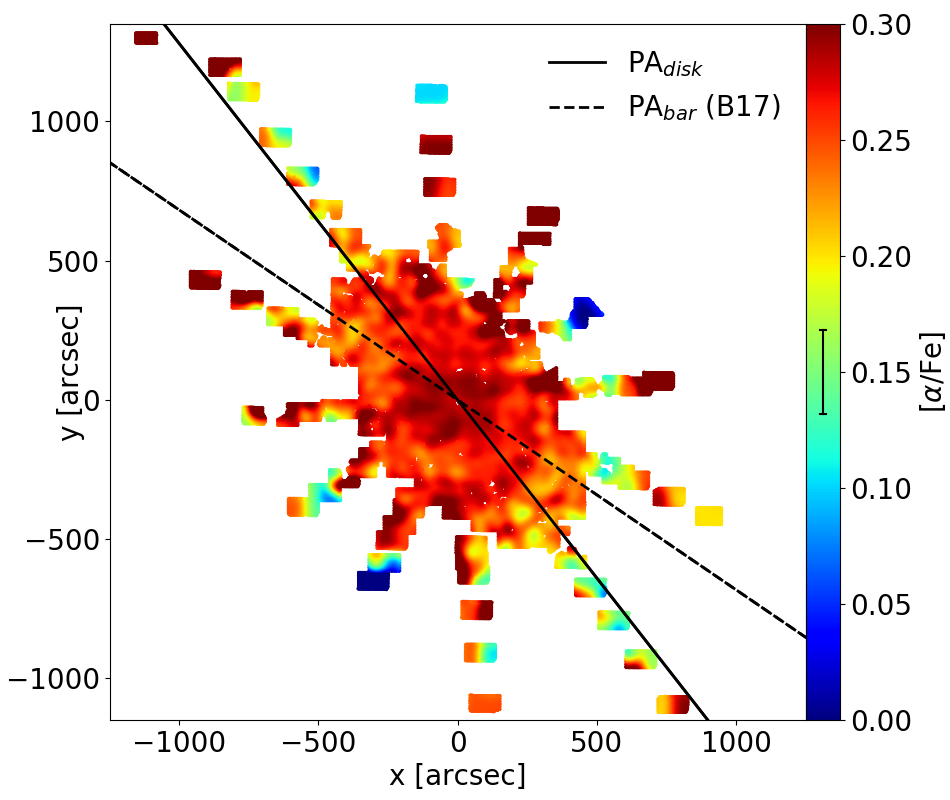}}
\caption[$\alpha/$Fe-overabundance map]{Mean [$\alpha/$Fe] overabundance map. The 
lines are as in Fig. \ref{fig:age}. See Figs. \ref{fig:schema} and \ref{fig:regions} for a schematic description of the geometry of the galaxy, its components, and the distribution of the dust.}
\label{fig:aFe}
\end{figure}

\begin{figure}
\resizebox{\hsize}{!}{\includegraphics{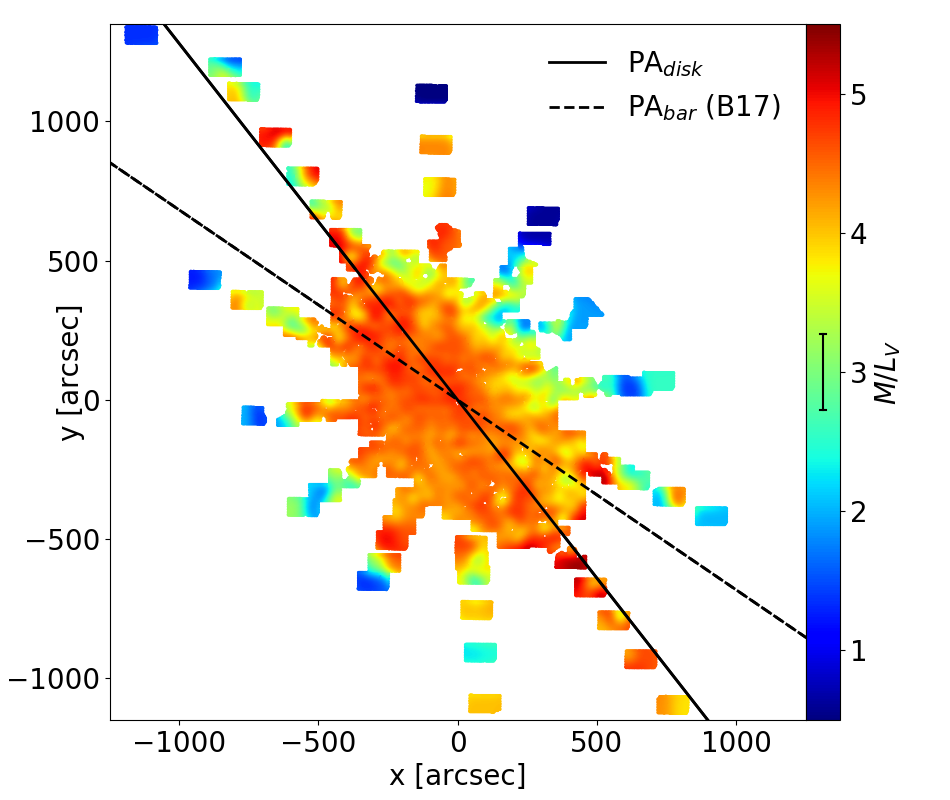}}
\caption[Mass to light ratio in V-band]{Mass-to-light ratio in V band.
The lines are as in Fig. \ref{fig:age}. See Figs. \ref{fig:schema} and \ref{fig:regions} for a schematic description of the geometry of the galaxy, its components, and the distribution of the dust.}
\label{fig:ML}
\end{figure}

\begin{figure}
\resizebox{\hsize}{!}{\includegraphics{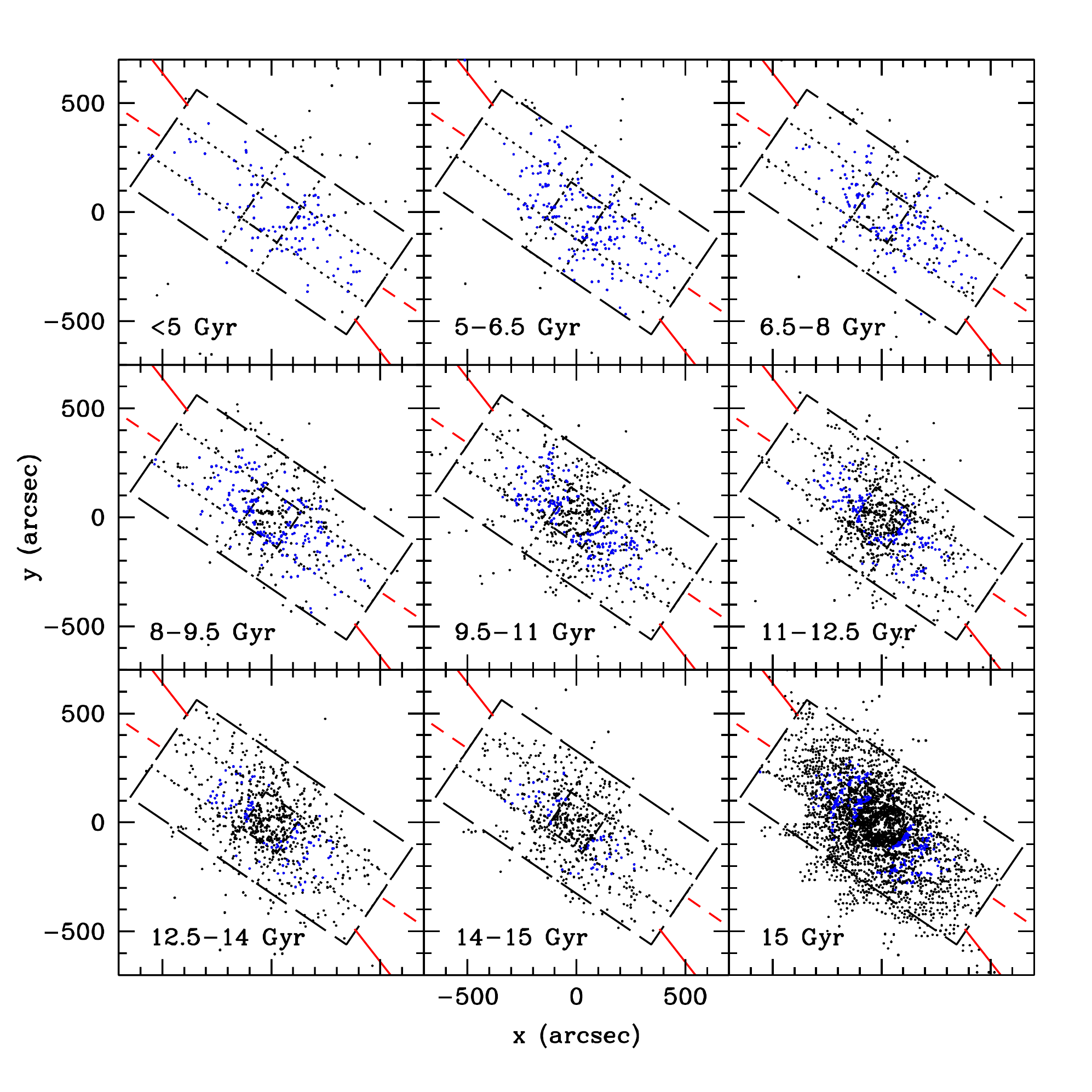}}
\caption[xy age distribution]{Positions on the sky of the analyzed
  spectra grouped by age bins. The positions of spectra sampling less
  than $10^6M_\odot$ are plotted in blue.}
\label{fig:agexy}
\end{figure}

The Figures \ref{fig:age}, \ref{fig:ZH}, \ref{fig:aFe}, and
\ref{fig:ML} show the spatial distributions (smoothed with a
  two-dimensional Gaussian filter with $\sigma=10$\arcsec) of the age,
   metallicity, $\alpha-$element overabundance, and $M/L_V$
  ratios.
The smooth appearance of Fig. \ref{fig:age} masks the spread of
  the spatial distribution of measured ages: Fig. \ref{fig:agexy} shows that
  at almost every position in the bulge region we observe a mixture
  of between 40 and 50\% of 15 Gyr values and the almost flat
  distribution of ages between 4 and 15 Gyr already seen in
  Fig. \ref{fig:distribution}. In contrast, in the disk region 80\%
  of the age values are lower than 15 Gyr. 

  \citet{Dong2018} analyzed the color-magnitude diagrams of the inner
  $\approx 200$ arcsec bulge region of M31, detecting a widespread
  population of stars younger than 5 Gyr. Some 80\% of the stellar
  mass outside 100 arcsec from the center should be older than 8 Gyr
  and metal rich ([Fe/H]$\sim0.3$). We note that for 80\% of our spectra
  we determine a Lick-based age that is larger than 10 Gyr (see
  Fig. \ref{fig:distribution}). Some 6\% of the total mass in this
  region should be younger than 2 Gyr.  Moreover, \citet{Dong2018}
  reported the existence of $\approx 1300- 5000$ bright (100-10000
  $L_\odot$) young ($\le 1$ Gyr) stars in the inner 100-200 arcsec of
  M31.  Our single fibers cover 8 arcsec$^2$, sampling of the order of
  $10^5 L_\odot$ at 100-200 arcsec from the center; in addition, these fibers
  cover one-third of the available area, every 3 or 1 out of 10 fibers contain one of these stars.

  Following the method described in Sect. \ref{sec:CBBP} (see
  Eq. \ref{eq:mixIndex}), we computed the Lick indices that a mix of
  SSPs distributed in age and mass as derived by
  \citet[][see their Fig. 19]{Dong2018} would have assuming
  [Z/H]=0.35 and [$\alpha/$Fe]=0.3; we note that the oldest population
  component considered there has an age of 12 Gyr. We compared the
  derived mean luminosity-weighted age to the true mass-weighted age
  of the mixture. We find that Lick ages larger than 8 Gyr can be
  obtained only in the absence of components younger than 3-4 Gyr. In
  this case Lick ages estimate the mass-weighted age within 0.1
  Gyr. Lick ages lower than 5-6 Gyr are measured if the 1-2 Gyr
  population component is present. In this case Lick ages
  underestimate the mass-weighted age by 1-2 Gyr and, as a
  consequence, the derived $M/L_V$ by up to 70\%. It is therefore
  plausible that the distribution of ages of
  Fig. \ref{fig:distribution} originates from a combination of
  measurement errors and the random sampling of a mixture of a
  majority of old ($\ge 8$ Gyr) stars with a minority of younger
  ($\le 4$ Gyr) stars.

  Finally, it is important to note that there is no difference between
  the stellar kinematics (velocity and velocity dispersions) of fibers
  tagged with an old ($\ge 5$ Gyr) or young ($<5$ Gyr) Lick
  age. Therefore the fibers tagged with a young age belong to the
  bulge region and do not sample disk stars seen in projection.

The mean metallicity similar to the solar value quoted by
\citet{Sarajedini05} is compatible with the values we derive for the
disk region (see Fig. \ref{fig:GradientDisk}). \citet{Gregersen15}
derived a metallicity using observed individual red giant branch
stars outside $R \approx$ 720\arcsec. They measure an enhancement of
metallicity in the bar region, in agreement with our findings, and a
decline of the metallicity outside 720\arcsec , which is not seen in
our data. However, they assume a constant age of 4 Gyr for their data,
which might not be appropriate (see Fig. \ref{fig:GradientDisk}).

Figs. \ref{fig:age_Saglia}, \ref{fig:ZH_Saglia}, and
\ref{fig:aFe_Saglia} in the Appendix compare our values based on the
\citet{Thomas03} models to those measured by \citetalias{Saglia10} and find an overall good agreement. Their young central population,
  coupled with a metallicity and an overabundance gradient, is
  somewhat washed out by the size of our fibers and the Gaussian
  filtering applied in this work, but can be seen in the
  Figs. \ref{fig:agexy}, \ref{fig:GradientMN}, and
  \ref{fig:GradientMJ}.


\begin{figure}
\resizebox{\hsize}{!}{\includegraphics{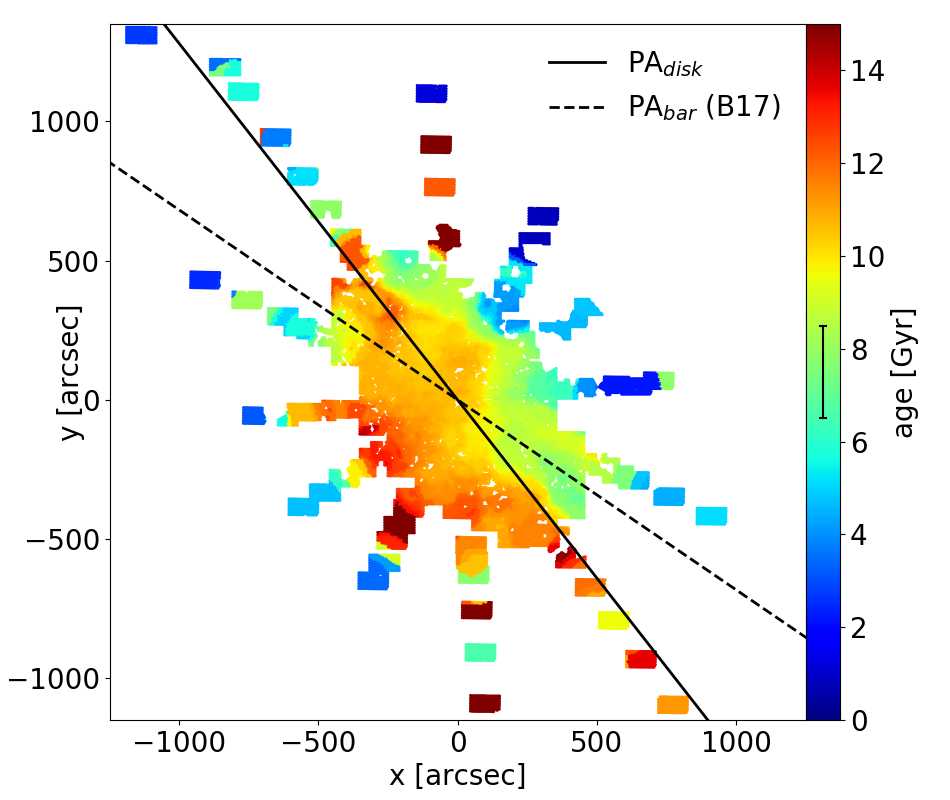}}
\caption[Age80 map]{Map of the ages corresponding to the 20\%
  percentile of the ages in a radius of 100 arcsec around each
  point. The lines are as in Fig. \ref{fig:LickHbeta_map}. See
  Figs. \ref{fig:schema} and \ref{fig:regions} for a schematic
  description of the geometry of the galaxy, its components, and the
  distribution of the dust. }
\label{fig:age80map}
\end{figure}

We turn now to a discussion of the mean age, metallicity,
  [$\alpha$/Fe] overabundance, and $M/L_V$ maps. As discussed above,
  the overall distributions of the metallicity,
  [$\alpha$/Fe] overabundance, and $M/L_V$ values are Gaussian-like and
  defining a meaningful spatially dependent mean is not a
  problem. This is not obviously true for the age values, where a
  large percentage of best-fitting ages hit the maximum allowed by the
  grid. Therefore, guided by the fact that 80\% of our measured ages
  are older than 10 Gyr (see Fig. \ref{fig:distribution}), we
  construct a second age map (Fig. \ref{fig:age80map}), where at each
  point we plot the age corresponding to the 20\% percentile of the
  ages in a radius of 100 arcsec around the point; i.e., 80\% of the
  measured population ages in the 100 arcsec radius are older than
  this value. We tabulate these values dubbed Age20\% in Table
  \ref{tab:Stellar_populations}, where we also report the $M/L_V$20\%
  we derive with these ages at the metallicity [Z/H] of each point.

Both this map and the mean age map of Fig. \ref{fig:age} are
reminiscent of the H$\beta$ map of Fig. \ref{fig:LickHbeta_map}; the
overall distribution is asymmetric and lower values in the upper
half are relative to the disk major axis and higher values in the lower
half. As discussed above, we relate this gradient to the dust present
in the disk of M31, that is masking the old stellar population of the
bulge in the upper half. On average, the two maps differ by 2.4
  Gyr, which can be seen as the systematic uncertainty of our age
  estimates. The most obvious systematic
  difference between the two maps, the 6 to 8 Gyr patch visible to the
  northwest of the center in Fig. \ref{fig:age80map}, is probably due
  to the 5 kpc star formation ring discussed by \citet[][see their
  Fig. 12]{Dong2018}.

\begin{figure}
\resizebox{\hsize}{!}{\includegraphics{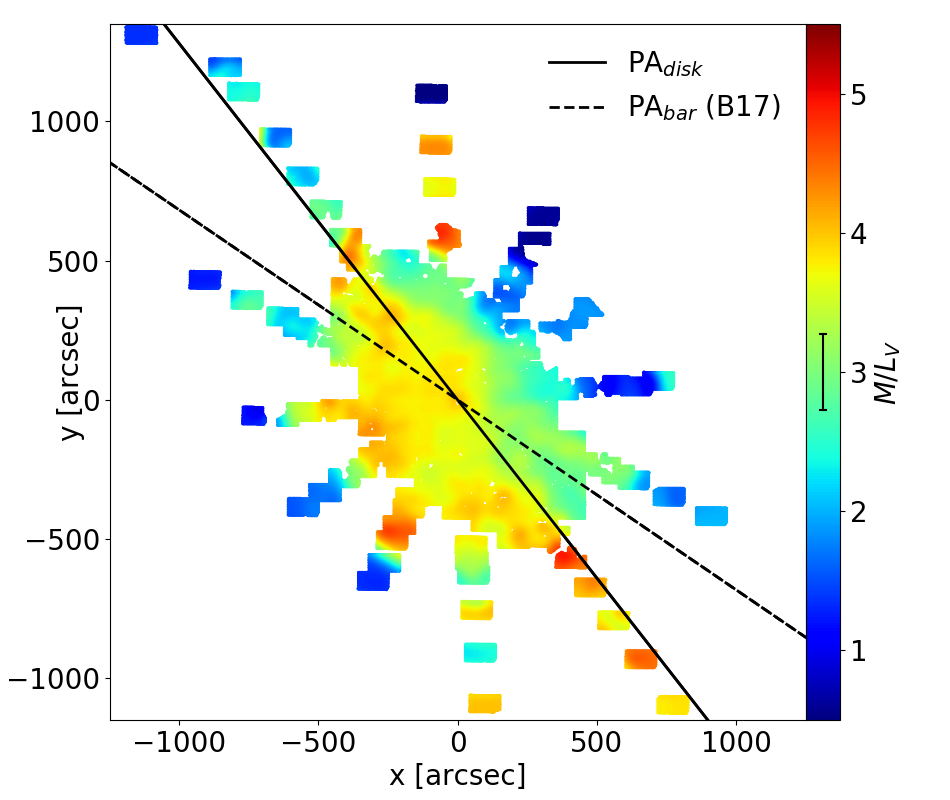}}
\caption[Age80 map]{Mass-to-light ratio in V band computed using the
  ages shown in Fig. \ref{fig:age80map} (see columns Age20\% and
  $M/L_V$20\% of Table \ref{tab:Stellar_populations} ). The lines are as in
  Fig. \ref{fig:age}. See Figs. \ref{fig:schema} and \ref{fig:regions}
  for a schematic description of the geometry of the galaxy, its
  components, and the distribution of the dust.}
\label{fig:MLVage80map}
\end{figure}

The mean metallicity in Fig. \ref{fig:ZH} is enhanced along the
bar position angle, such as the Mg~b and iron indices.  The mean [$\alpha$/Fe]-overabundance map of Fig. \ref{fig:aFe} shows
that [$\alpha$/Fe] is enhanced in the central region, but otherwise
relatively constant over the whole area without the bar signature of
the metallicity.  The mean $M/L_V$ map of Fig \ref{fig:ML} looks
similar to the age map and has lower values in the top right and
higher values in the bottom left.  Fig. \ref{fig:MLVage80map} shows
  the map of the $M/L_V$20\% values reported in Table
  \ref{tab:Stellar_populations} (smoothed with a two-dimensional
  Gaussian filter with $\sigma=10$\arcsec). They are on average 0.8
  $M_\odot/L_\odot$ smaller than the $M/L_V$ given there. The map
  mirrors closely the topology observed in Fig. \ref{fig:age80map}.

\section{Discussion}
\label{sec:Discussion}

In the following, we concentrate on interpreting the mean
  trends mentioned above.

\subsection{Is there a young disk component in the bulge region?}
\label{sec:BD}
 
The B17 model of M31 produces a B/P bulge component through a bar
buckling phase that destroys the original old primordial disk.  The
age maps of Figs. \ref{fig:age} and \ref{fig:age80map} show that the outer
disk regions of M31 can be as young as 4 Gyr. In Paper I we used the
major axis photometric decomposition of \citet{Kormendy99} shown
  in Fig. \ref{fig:Kormendy} to define the bulge and disk region
of the galaxy. In this paper, we want to investigate whether a cold, rotating
disk component in the inner 600 arcsec region of M31 is compatible
with the observed kinematics and whether it could be as young as the
outer disk (4 Gyr). We note, however, that \citet{Kormendy99} did not
claim that their decomposition implies the presence of such a disk.

First, we assess whether the line-of-sight velocity distributions
(LOSVDs) $\mathcal{F}(v)$ presented in Paper I can be explained by the
sum of two Gaussians $\mathcal{B}(v_{bulge},\sigma_{bulge})$ and
$\mathcal{D}(v_{disk},\sigma_{disk})$, describing the kinematics of a
bulge and a disk component with the bulge-to-total fraction
$\frac{B}{T}$ and the disk-to-total fraction $\frac{D}{T}$ values
taken from the model image of \citetalias{Opitsch18} as follows:
\begin{equation}
\mathcal{F}(v)= \frac{B}{T} \cdot \mathcal{B}(v_{bulge},\sigma_{bulge}) + \frac{D}{T} \cdot \mathcal{D}(v_{disk},\sigma_{disk}).
\label{eq:Fv}
\end{equation}
Fitting an (almost) Gaussian LOSVD with two Gaussians is inherently
degenerate and at least two almost equivalent solutions can be
found. One of these is a hot bulge with little rotation plus a colder,
quickly rotating disk. A second solution produces a bulge with
cylindrical rotation and a disk with on average zero rotational
velocity in the inner 100 arcsec.  Performing the decomposition
described by Eq. \ref{eq:Fv} on the models of B17 yields qualitatively
the same results with similar solutions.

As a second step, we build a two-component spectral model. We choose
one spectrum from the very center of M31 as the bulge spectrum and the
outermost bin along direction PA=55$^{\circ}$ on the eastern side as
the disk spectrum. We fit both the bulge and disk spectrum with
\texttt{pPXF} to obtain the optical linear combinations of the
template star spectra that best describe them.  For these fits, all
the spectra of the 230 giants from the \texttt{ELODIE} catalog
\citep{Prugniel07} are used and broadened to the \texttt{VIRUS-W}
resolution. The fits result in a model bulge and a model disk
spectrum. We broaden the bulge and disk model spectra with the LOSVDs
$\mathcal{B}(v_{bulge},\sigma_{bulge})$ and
$\mathcal{D}(v_{disk},\sigma_{disk})$ determined above, considering
both kinematic solutions, and sum the broadened model spectra. We
measure the Lick indices on these combined model spectra with the
method described in Sec. \ref{sec:StellarPopulations}. We then fit
age, metallicity, and [$\alpha/$Fe] overabundance in a similar way as
in Sect.  \ref{sec:StellarPopulations}. The resulting age map
\citep[see Fig. 6.35 in][]{Opitsch2016} shows a strong radial gradient
from the center, where the old population of the bulge spectrum
dominates, to the outer regions, where the spectrum of the young disk
population becomes important. This is not compatible with what we
observe in Fig. \ref{fig:age} or \ref{fig:age80map}. The result holds
whatever kinematic decomposition we use. We conclude that a young (4
Gyr) disk component distributed as the photometric decomposition of
\citet{Kormendy99} is not present in the inner 600 arcsec region of
M31. This result is not surprising, given that on average 80\% of our
age measurements are older than 10 Gyr, while the exponential
component of \citet{Kormendy99} contributes more than 20\% of the
light at distances larger than 200 arcsec from the center.  We could
accommodate the mild age gradient along the major axis of M31 (see
Fig. \ref{fig:GradientMJ}, top left) if the disk component distributed
as the photometric decomposition of \citet{Kormendy99} were as old as
the bulge at the center and became progressively younger with
distance, reaching 6 Gyr at 600 arcsec.

As discussed in the Introduction, \citet{Blana18} presented a model
  of the inner region of M31 as the sum of a bar plus B/P
  component and a CB. Fig. \ref{fig:Kormendy} shows
  their profiles along the bar region and along the position angle of
  the major axis of M31, as V-band luminosity densities projected
  along the line of sight, taking into account dust attenuation. The
  comparison with the decomposition of \citet{Kormendy99} reveals the
  following.  By construction, the B/P component matches the profile
  of the disk outside the bar region ($R>600$ arcsec), in agreement
  with \citet{Kormendy99}. However, inside this region much of what
  \citet{Kormendy99} referred to as the bulge component is, in reality, the bar
  plus B/P structure. As a consequence, the CB component
  of \citet{Blana18} model has a much shorter scale
  length than that of \citet{Kormendy99}.

\begin{figure}
\resizebox{\hsize}{!}{\includegraphics{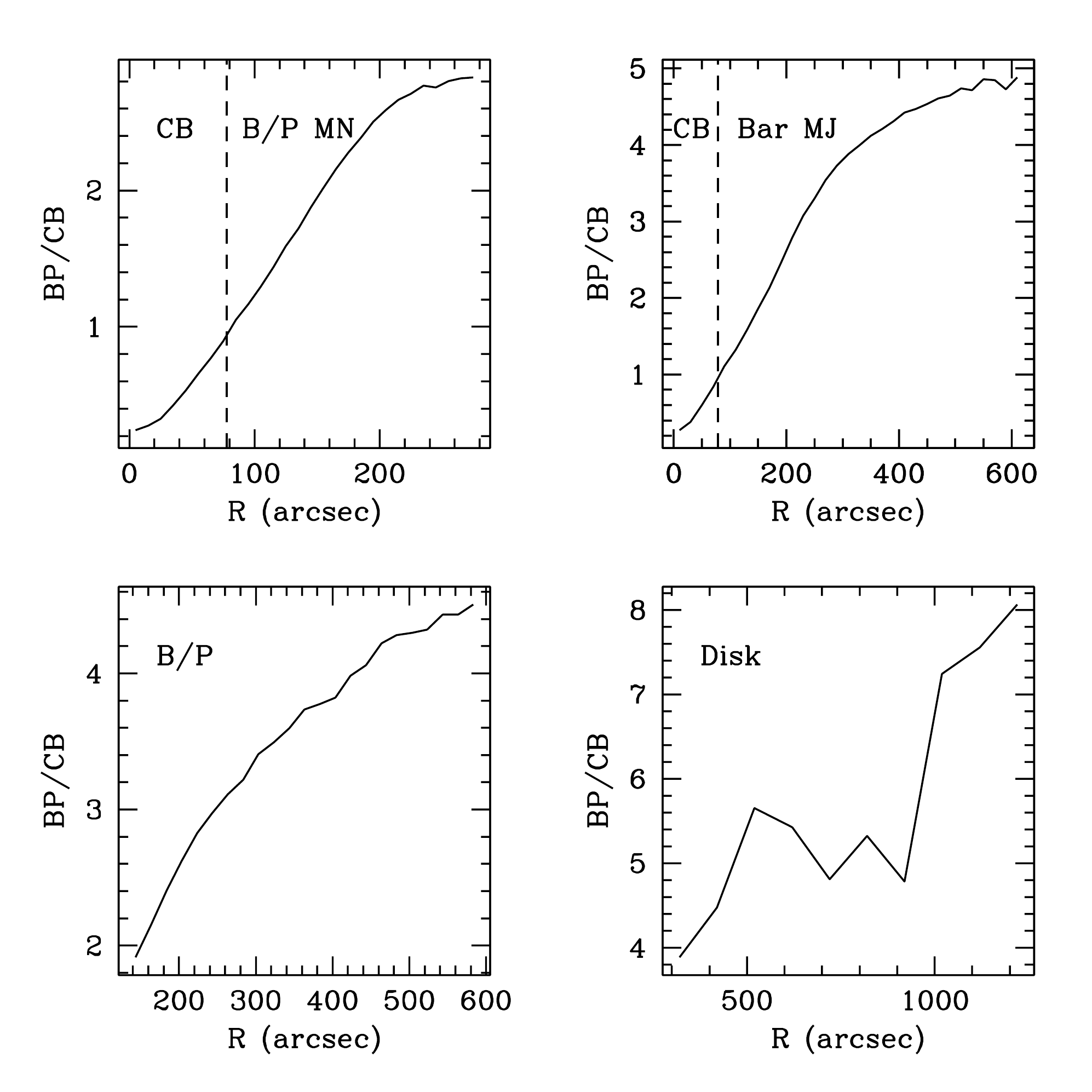}}
\caption[Gradients along the MJ axis of the bar] {Ratios BP/CB
  between the projected V-band luminosity density of the B/P and CB components of the model of
\citet{Blana18} for the four regions shown in
  Fig. \ref{fig:regions}. }
\label{fig:DB}
\end{figure}

  Fig. \ref{fig:DB} shows the ratios BP/CB between the projected
  V-band luminosity densities, taking into account dust attenuation,
  of these two components for the four regions shown in
  Fig. \ref{fig:regions}. At distances smaller than 100 arcsec from
  the center the ratio BP/CB is less than one and CB dominates the
  V-band light.  In contrast, the ratio BP/CB is always larger than 2
  in the B/P bulge (denoted in green in Fig. \ref{fig:regions}) and
  larger than 4 in the disk (denoted in cyan in Fig. \ref{fig:regions})
  regions. We make use of these BP/CB profiles in the next
  section.

\subsection{Classical and B/P bulge, bar, and disk}
\label{sec:CBBP}

\begin{figure*}
\resizebox{\hsize}{!}{\includegraphics{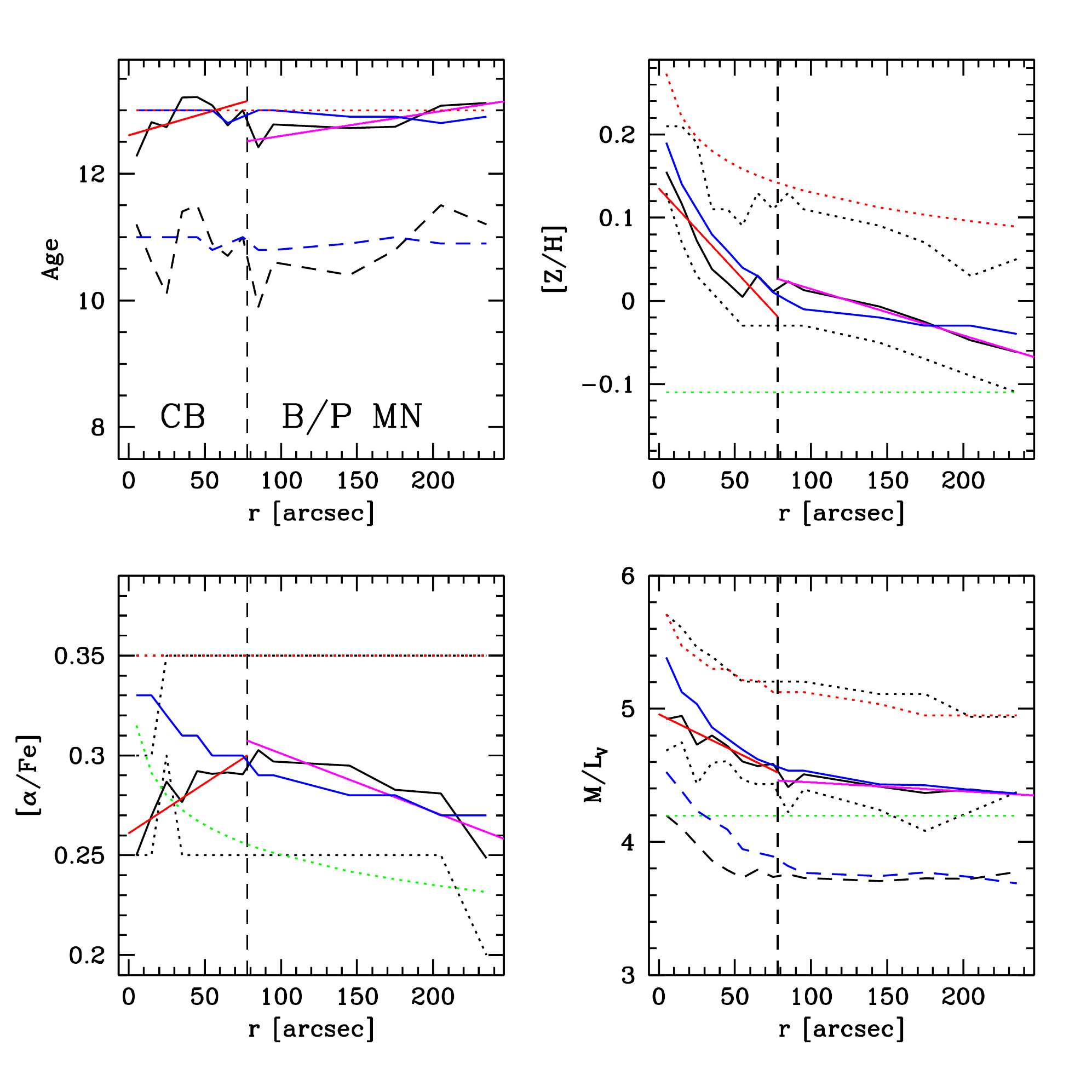}}
\caption[Gradients along the MN axis of the bar]{Average (over
  regions 10 arcsec wide in the inner 100 arcsec, 30 arcsec wide at
  larger distance) age, metallicity, [$\alpha$/Fe], and $M/L_V$
  profiles along the bar minor axis of M31 (black full lines)
  tracing the CB component of M31 (see
  Fig. \ref{fig:regions}). The black dashed line in the $M/L_V$
    plot shows the average (over the same regions) of the $M/L_V$20\%
    values. The black dashed line in the age plot shows the 20\%
    percentile of the distribution of values in the bins; the black
    dotted lines in the other plots show the 34 and 68 percentile of the distributions of
    values in the bins.  The straight lines (red for the inner bulge,
  magenta for the outer bulge; see Fig. \ref{fig:regions}) show the
  fitted gradients (see Table \ref{tab:gradients}).   The blue
    lines show the composite stellar population model according to
    Eq. \ref{eq:mixIndex} and Table \ref{tab:ParamDB}, the dotted red
    and green lines the assumed profiles for the CB and B/P
    components, respectively. The blue dashed lines show the age and
    $M/L_V$ profiles obtained using the age values denoted with
    ``20\%'' in Table \ref{tab:ParamDB}.  The vertical dashed line
  indicates 0.13$r_{bar}$.  }
\label{fig:GradientMN}
\end{figure*}

\begin{figure*}
\resizebox{\hsize}{!}{\includegraphics{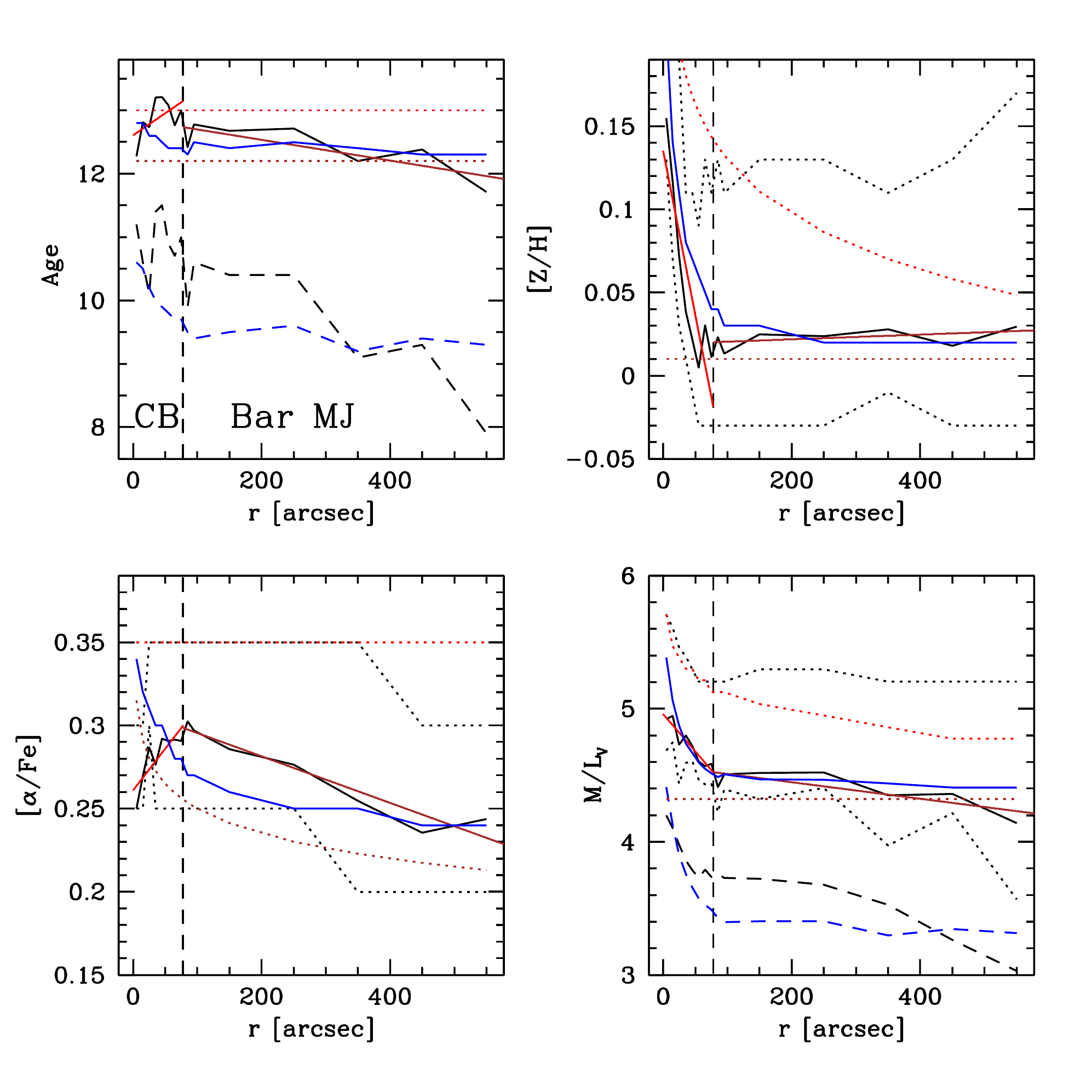}}
\caption[Gradients along the MJ axis of the bar] {Average (over
  regions 10 arcsec wide in the inner 100 arcsec, 100 arcsec wide at
  larger distance) age, metallicity, [$\alpha$/Fe], and $M/L_V$
  profiles along the bar major axis of M31 (black full lines),
  tracing the bar of M31 (see Fig. \ref{fig:regions}).  The black
    dashed line in the $M/L_V$ plot shows the average (over the same
    regions) of the $M/L_V$20\% values. The black dashed line in the
    age plot shows the 20\% percentile of the distribution of values
    in the bins; the black dotted lines in the other plots show the 34 and 68
    percentile of the distributions of values in the bins. The
  straight lines (red for the inner bulge, brown for the bar, see
  Fig. \ref{fig:regions}) show the fitted gradients (see Table
  \ref{tab:gradients}). The blue lines show the composite stellar
    population model according to Eq. \ref{eq:mixIndex} and Table
    \ref{tab:ParamDB}, the dotted red and brown lines the assumed
    profiles for the CB and bar, respectively. The blue dashed
    lines show the age  and
    $M/L_V$ profiles obtained using the age values denoted
    with ``20\%'' in Table \ref{tab:ParamDB}. The vertical dashed
  line indicates 0.13$r_{bar}$ (see text). }
\label{fig:GradientMJ}
\end{figure*}

\begin{figure}
\resizebox{\hsize}{!}{\includegraphics{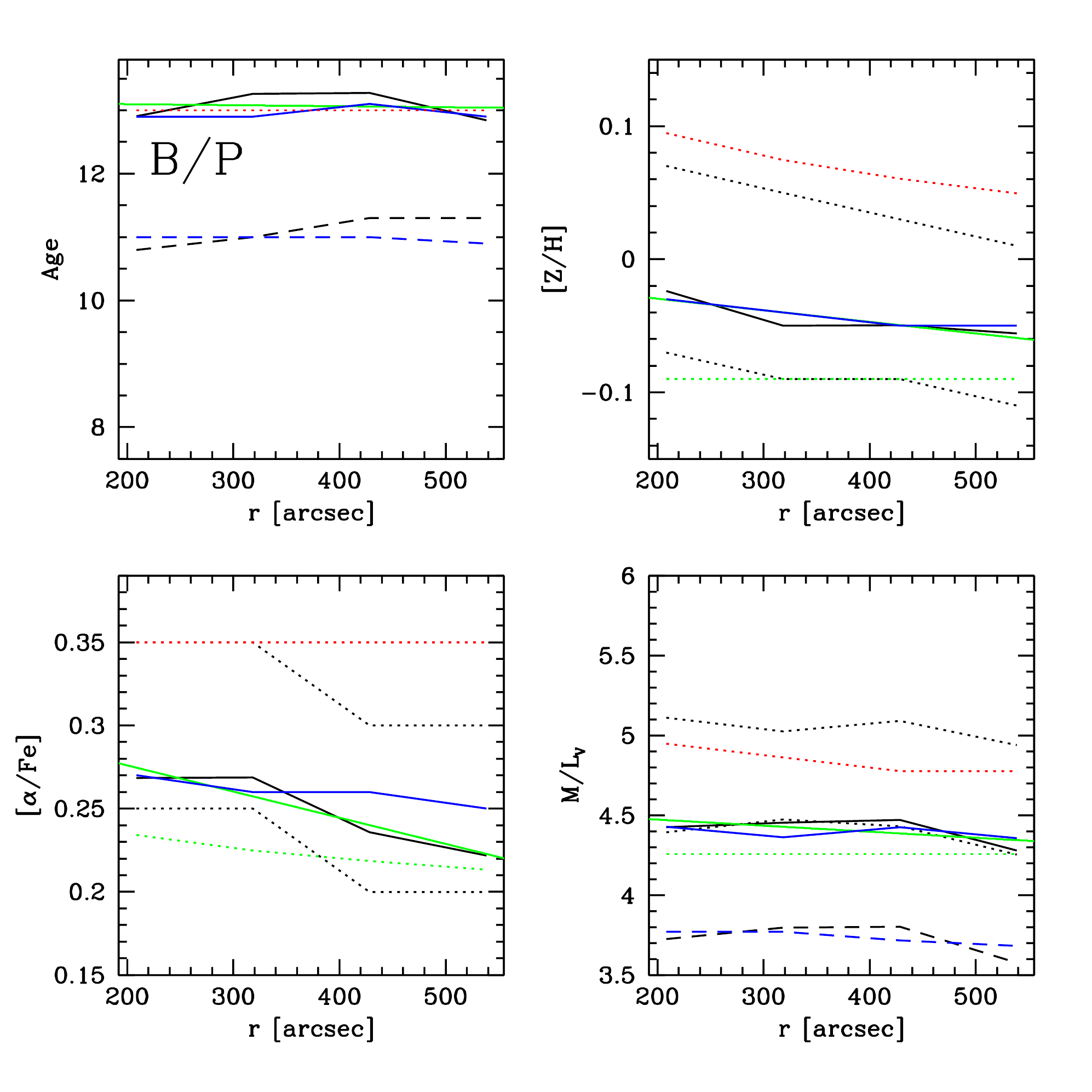}}
\caption[Gradients along the MN axis of the bar]{Average (over
  regions 110 arcsec wide in radial distance) age, metallicity,
  [$\alpha$/Fe], and $M/L_V$ profiles in the B/P bulge region of M31
  (black full lines; see Fig. \ref{fig:regions}).  The
    black dashed line in the $M/L_V$ plot shows the average (over the
    same regions) of the $M/L_V$20\% values. The black dashed line in
    the age plot shows the 20\% percentile of the distribution of
    values in the bins; the dotted lines in the other plots show the 34 and 68
    percentile of the distributions of values in the bins. The
  straight green (see Fig. \ref{fig:regions}) line shows the fitted
  gradient (see Table \ref{tab:gradients}). The blue lines show
    the composite stellar population model according to
    Eq. \ref{eq:mixIndex} and Table \ref{tab:ParamDB}, the dotted red
    and green lines the assumed profiles for the CB andB/P
    components, respectively. The blue dashed lines show the age and
    $M/L_V$ profiles obtained using the age values denoted with
    ``20\%'' in Table \ref{tab:ParamDB}.} 
\label{fig:GradientPseudo}
\end{figure}

\begin{figure}
\resizebox{\hsize}{!}{\includegraphics{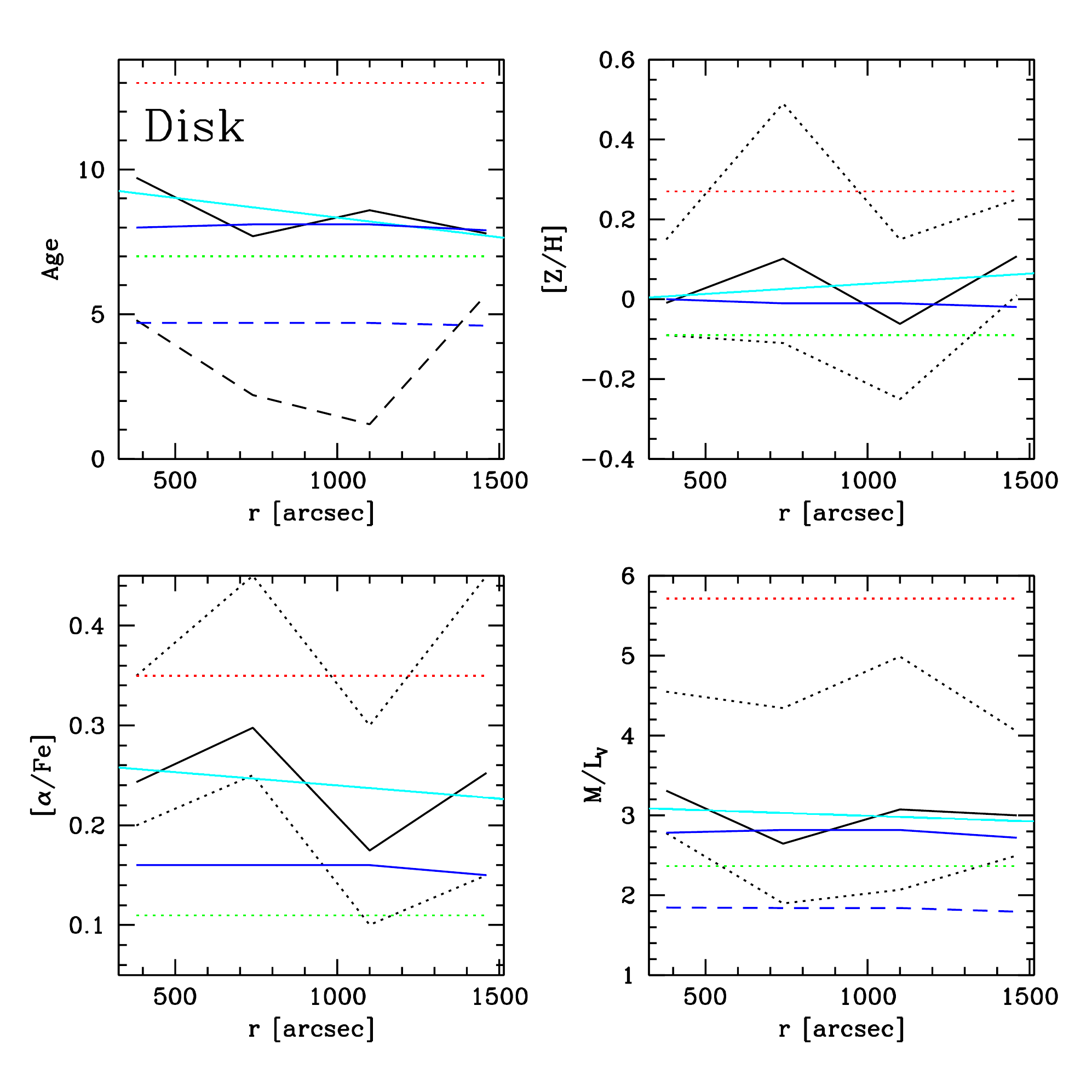}}
\caption[Gradients along the MN axis of the bar]{Average (over
  regions 360 arcsec wide in radial distance) age, metallicity,
  [$\alpha$/Fe], and $M/L_V$ profiles in the disk region of M31 (
    (black full lines), see Fig. \ref{fig:regions}).  The black
    dashed line in the $M/L_V$ plot shows the average (over the same
    regions) of the $M/L_V$20\% values.  The black dashed line in the
    age plot shows the 20\% percentile of the distribution of values
    in the bins; the black dotted lines in the other plots show the 34 and 68
    percentile of the distributions of values in the bins. The
  straight cyan (see Fig. \ref{fig:regions}) line shows the fitted
  gradient (see Table \ref{tab:gradients}).  The blue lines show
    the composite stellar population model according to
    Eq. \ref{eq:mixIndex} and Table \ref{tab:ParamDB}; the dotted red
    and green lines the assumed profiles for the CB and B/P
    components, respectively. The blue dashed lines show the age and
    $M/L_V$ profiles obtained using the age values denoted with
    ``20\%'' in Table \ref{tab:ParamDB}.}
\label{fig:GradientDisk}
\end{figure}

In order to characterize the mean stellar populations of the
classical and B/P bulge, bar, and disk components, we computed profiles,
averaged over radial bins chosen to smooth out the spread of the
  single measurements, along four directions, plotting these as a
function of the distance from the center. As for the age map, we
  also provide the profiles of the ages corresponding to the 20\%
  percentiles of the age distributions in the bins.
  Fig. \ref{fig:GradientMN} shows the values of the age, metallicity,
  [$\alpha$/Fe], and $M/L_V$ obtained by averaging the age,
  metallicity, [$\alpha$/Fe], and $M/L_V$ in radial bins (10 arcsec
  wide in the inner 100 arcsec, 30 arcsec wide at larger distance)
  over a rectangular region 200 arcsec wide and 450 arcsec long in the
  CB and minor axis region (perpendicular to the bar) and taking into
  account only points with low absorption ($R_{dust}>0.8$, see
  Fig. \ref{fig:regions}).  Fig. \ref{fig:GradientMJ} shows the same
  profiles in the bar region, averaged in radial bins (10 arcsec wide
  in the inner 100 arcsec, 100 arcsec wide at larger distance) over a
  rectangular region 200 arcsec wide and 1200 arcsec along the bar at
  $PA=55.7$ deg, again with $R_{dust}>0.8$ (see
  Fig. \ref{fig:regions}).  Fig. \ref{fig:GradientPseudo} presents
the profiles obtained in the B/P bulge region with low absorption (see
Fig. \ref{fig:regions}) using radial bins 110 arcsec wide.  Finally,
Fig. \ref{fig:GradientDisk} shows the profiles for the disk region,
outside the B/P bulge in the northern, dusty part of M31 (see
Fig. \ref{fig:regions}) using radial bins 360 arcsec wide. In Table
\ref{tab:mean_population_measurements} we give the average values and
rms of these profiles and in Table \ref{tab:gradients} the
linear gradients and zero points measured from the profiles shown in
Figs. \ref{fig:GradientMN} to \ref{fig:GradientDisk}.

As discussed above, capping the maximum ages to the age of the
  universe shifts the mean ages and profiles of the bar, CB and B/P
  regions by approximately 0.6 Gyr and by 0.3 Gyr in the disk. The age
  gradients quoted in Table \ref{tab:gradients} vary less than the
  quoted errors.

  Table \ref{tab:Age80} reports the means and rms, gradients, and
  zero points of the age profiles shown by the black dotted lines
  (based on the 20\% percentile of the distribution of the age values
  in the bins) in the Figs. \ref{fig:GradientMN} to
  \ref{fig:GradientDisk}. In this case the ages of the bar, CB, and B/P
  regions are approximately 2 Gyr younger and the disk appears 5 Gyr
  younger.

In order to
  partially address the fact that we are using SSPs to interpret the
  line strengths of a mix of stellar populations, we construct the
  following simple two-populations model. We assume that the CB
  (marked orange and magenta in the top left and right plots of
  Fig. \ref{fig:regions}) is made of stars of the same age $Age_{CB}$
  and $\alpha-$element overabundance [$\alpha/$Fe]$_{CB}$ with a
  metallicity profile written as
\begin{equation}
\label{eq:ZH_CB}
[Z/H]_{CB}(R)=a_{CB}\log R/arcsec +[Z/H]_{CB}(0).
\end{equation}
Moreover, the bar (indicated in brown in the top right plot of
Fig. \ref{fig:regions}), the B/P bulge (indicated in green in the bottom left
plot of Fig. \ref{fig:regions}) and the disk (indicated in cyan in the bottom
right plot of Fig. \ref{fig:regions}) are described by stellar
populations with constant age $Age_{Bar}$, $Age_{B/P}$, $Age_{Disk}$,
metallicity $[Z/H]_{Bar}$, $[Z/H]_{B/P}$, $[Z/H]_{Disk}$, and profile
of $\alpha-$element overabundance 
\begin{equation}
\label{eq:aFe_BP}
[\alpha/Fe]_{B/P}(R)=a_{B/P}\log R/arcsec +[\alpha/Z]_{B/P}(0).
\end{equation}
At each radius and for each region of Fig. \ref{fig:regions} we
compute the line strength indices $Index_{CB}$ and $Index_{Region}$ (where 
$Region$ indicates $B/P$, $Bar,$ or $Disk$, according to the region considered)
appropriate for the age, metallicity, and $\alpha-$overabundance
corresponding to the assumptions made above, and we derive the indices
$Index$ of the sum of the components as%
\begin{equation}
\label{eq:mixIndex}
Index(R)= \frac{Index_{CB}(R)+BP/CB(R)\times Index_{Region}(R)}{1+BP/CB(R)}.
\end{equation}
Finally, we compute the age, metallicity, and $\alpha-$overabundance of
the SSP that best reproduce the set of indices at each radius by
minimizing Eq. \ref{eq:chi}. Table \ref{tab:ParamDB} reports the
values of the parameters chosen to produce the profiles shown by the
blue lines in Figs. \ref{fig:GradientMN}-\ref{fig:GradientDisk}. The age profiles shown as blue dotted lines in
the same figures are computed using the values denoted as ``20\%'' in
the table. The radial variations of these profiles are the result of
the combination of the varying radial contributions of the CB, bar, and
B/P components captured by the $BP/CB(R)$ function and the intrinsic
gradients modeled by Eqs. \ref{eq:ZH_CB} and \ref{eq:aFe_BP}. The
proposed solution is by no means unique, but its existence makes the
following arguments plausible. 

We start with a discussion of the CB region, which
  dominates the inner 78 arcsec (see Fig. \ref{fig:DB}) and can be
  characterized best by focusing on the region orthogonal to the bar
  shown in Fig.  \ref{fig:GradientMN}.  The age profile is essentially
  constant, which we model assuming $Age_{CB}=Age_{B/P}$ in Table
  \ref{tab:ParamDB}. The central decrease in age is the low-resolution
  appearance of the central young population detected in the high-resolution data of \citet[][their Fig. 12]{Saglia10}. 

  The metallicity of the CB region is high at the center
  ([Z/H]$=0.14\pm0.02$) and declines outward in broad agreement with 
\citet{Dong2018} .  There is a change in
  slope of the metallicity gradient at roughly 78 arcsec, or 0.13
  $r_{bar}$, with $r_{bar}$=600\arcsec \ taken from B17, from rather
  steep (-0.5 dex/kpc, entry `InBulge' in Table \ref{tab:gradients},
  corresponding to the orange region of Fig. \ref{fig:regions}), to
  -0.1 dex/kpc (entry `OutBulge' in Table \ref{tab:gradients},
  corresponding to the magenta region in Fig.\ref{fig:regions}).
  Similar gradients and changes of slopes at this radius are measured
  in barless galaxies or along the minor axis of barred galaxies
  \citep{Seidel16}.  Our simple model captures this trend with
  Eq. \ref{eq:ZH_CB}.

  We modeled the complex behavior of the $\alpha-$element profile with
  a constant [$\alpha/$Fe] value of the CB and the mild gradient
  (Eq. \ref{eq:aFe_BP} and Table \ref{tab:ParamDB}) for the bar and
  B/P components. The hint of a central decrease of [$\alpha/$Fe] is
  also seen in the higher resolution data of \citet[][their
  Fig. 12]{Saglia10} and is related to the recent episode of star
  formation discussed there. The central values of the age, [Z/H], and
  [$\alpha/$Fe] that we measure for M31 are similar to what
  \citet{Thomas06} compiled for CBs of similar velocity
  dispersion. The $M/L_V$ profile increases toward the center ($< 78$
  arcsec) as a result of the metallicity gradient of the CB and
  flattens out at larger distances. 

The cut along the bar of Fig. \ref{fig:GradientMJ} (top left) shows
that the bar (outside 78 arcsec from the center, corresponding to the
brown region in Fig. \ref{fig:regions}) is mainly made of old stars
($12.4\pm 0.6$ Gyr, entry `Bar' in Table
\ref{tab:mean_population_measurements}, or $9.7\pm1.0$ Gyr from
  Table \ref{tab:Age80}) 1-2 Gyr younger than those of the
CB (entry `InBulge' in the same tables). The (mild)
  observed age and $M/L_V$ gradients come into place through the
  mixing of the CB and bar populations. There is no metallicity
  gradient along the bar, but we need to assume the presence of a
  shallow negative [$\alpha$/Fe] gradient along the bar to explain the
  observed [$\alpha$/Fe] profile. These considerations are mirrored by
  the choice of parameters made in Table \ref{tab:ParamDB}. The
absence of a metallicity gradient along the bar is compatible with the small
positive gradients measured by \citet{Seidel16} for a sample of 16
large barred galaxies.

The stellar population in the B/P bulge region
(Fig. \ref{fig:GradientPseudo}) is old ($\approx 12.4\pm 0.4$ Gyr,
Table \ref{tab:mean_population_measurements}, entry `B/P', or
$11.1\pm 0.2$ Gyr from Table \ref{tab:Age80}) and has (sub)solar
metallicity ($\approx -0.04\pm 0.03$ dex) and overabundant in
$\alpha-$elements ($\approx 0.25\pm 0.02$ dex). This population also has  observed
  mild negative gradients that we explain through the effects of the
  declining metallicity of the CB (Eq. \ref{eq:ZH_CB}) and the
  gradient in the $\alpha-$element overabundance of the B/P component
  (Eq. \ref{eq:aFe_BP}).  Given the high inclination of M31, a
detailed three-dimensional model of the bar and B/P bulge region 
  with metallicity tags \citep{Portail17b} are needed to decide
whether the kinematic fractionation model of \citet{Debattista17} has
played a role in shaping the metallicity map we
observe. Interestingly, the hint of the pinching in the iron maps
noted above is also mirrored into the [Fe/H] map. The mild
  decline of $M/L_V$ profile is driven by the declining metallicity of
  the CB component.

Given the old ages and relatively uniform metallicity determined for
the three components (bar, classical, and B/P bulge) of the inner
600 arcsec of M31, it is not surprising that all of these have similar
$M/L_V$ ratios. Therefore we conclude that the underlying assumption
of a constant M/L ratio of the dynamical model presented in
B17 and \citet{Blana18} is valid.

The profiles derived in the disk region (Fig. \ref{fig:GradientDisk})
are more irregular, probably because of the presence of patchy dust
that also explain the irregular $BP/CB(R)$ profile of
  Fig. \ref{fig:DB} (bottom right plot).  The stars can be as young
as 3-4 Gyr, on average $\approx 8.7\pm 3.3$ Gyr
 (Table \ref{tab:mean_population_measurements}, entry
`Disk', or 3.5$\pm$2.1 Gyr from Table \ref{tab:Age80}),
have approximately solar
metallicity, are [$\alpha/$Fe] overabunant ($\approx 0.24\pm0.05$ dex)
and have low $M/L_V$ ($\approx 3\pm0.1 M_\odot/L_\odot$). We
  need to assume in our two-populations model that in this outer region the
  age of the B/P in the model of \citet{Blana18} is 
  much smaller than previously assumed (7 or 5 Gyr, see Table 
\ref{tab:ParamDB}) to  reproduce the observed
  values of the age and $M/L_V$.

\begin{table*}
  \caption[Mean Age, metallicity and aFe]{Mean value and rms scatter of the age, metallicity, 
[$\alpha/$Fe], and $M/L_V$ profiles shown in Figs. \ref{fig:GradientMN}
to \ref{fig:GradientDisk}. The different regions are defined in the text 
and Fig. \ref{fig:regions}.}
\label{tab:mean_population_measurements}
\begin{tabular}{l c c c c c c c c}
\hline
\hline
Region & Mean age & rms age & Mean [Z/H] & rms [Z/H] & Mean [$\alpha/$Fe] & rms [$\alpha/$Fe] & mean $M/L_V$ & rms $M/L_V$ \\
       & [Gyr]    & [Gyr]    & [dex]      & [dex]      & [dex]              &  [dex]             & [$M_\odot/L_\odot$] &  [$M_\odot/L_\odot$] \\
\hline
\hline
     InBulge           &       12.9           &      0.3           &     0.06           &     0.05           &      0.28           &    0.01           &       4.7           &      0.2          \\
    OutBulge           &       12.8           &      0.3           &    -0.02           &     0.03           &      0.28           &    0.02           &       4.4           &      0.1          \\
         Bar           &       12.4           &      0.4           &     0.02           &     0.01           &      0.27           &    0.02           &       4.4           &      0.1          \\
      B/P           &       13.1           &      0.2           &    -0.04           &     0.01           &      0.25           &    0.02           &       4.4           &      0.1          \\
        Disk           &       8.5            &      0.9           &     0.03           &     0.08           &      0.24           &    0.05           &       3             &      0.1          \\

\hline
\hline
\end{tabular}
\end{table*}

\begin{table*}
 \caption[Age, metallicity and aFe]{Gradients and zero points of  the age, metallicity, [$\alpha/$Fe], and $M/L_V$ profiles shown in Figs. \ref{fig:GradientMN}
 to \ref{fig:GradientDisk}. The different regions are defined in the text and Fig. \ref{fig:regions}.}
\label{tab:gradients}
\begin{tabular}{l c c c c c c c c}
\hline
\hline
Region & Age grad. & Age(r=0) & [Z/H] grad. & [Z/H](r=0) & [$\alpha/$Fe] grad. &  [$\alpha/$]Fe(r=0) & $M/L_V$ grad. & $M/L_V$(r=0) \\
       & [Gyr/kpc]    & [Gyr]    & [dex/kpc]      & [dex]      & [dex/kpc]              &  [dex]             & [$M_\odot/L_\odot$/kpc] &  [$M_\odot/L_\odot$] \\
\hline
\hline
     InBulge           &       1.7       $\pm$      1.0       &       12.6       $\pm$       0.2           &     -0.5       $\pm$    0.1            &     0.14       $\pm$    0.02           &     0.13       $\pm$     0.04           &      0.26       $\pm$     0.01           &      -1.5       $\pm$      0.3           &       5.0       $\pm$      0.1          \\
    OutBulge           &       1.0       $\pm$      0.3       &       12.2       $\pm$       0.2           &     -0.1       $\pm$    0.01           &     0.06       $\pm$    0.01           &    -0.07       $\pm$     0.02           &      0.32       $\pm$     0.01           &      -0.1       $\pm$      0.1           &       4.5       $\pm$      0.1          \\
         Bar           &      -0.4       $\pm$      0.2       &       12.9       $\pm$       0.2           &      0         $\pm$    0.01           &     0.02       $\pm$    0.01           &    -0.04       $\pm$     0.01           &      0.31       $\pm$     0.01           &      -0.2       $\pm$      0.05           &      4.6       $\pm$      0.1          \\
      B/P           &      -0.1       $\pm$      0.3       &       13.1       $\pm$       0.4           &      0         $\pm$    0.01           &    -0.01       $\pm$    0.13           &    -0.04       $\pm$     0.01           &      0.31       $\pm$     0.02           &      -0.1       $\pm$      0.09           &      4.5       $\pm$      0.1          \\
        Disk           &      -0.4       $\pm$      0.3       &       9.7        $\pm$       1.1           &      0.        $\pm$    0.03           &    -0.01       $\pm$    0.12           &    -0.01       $\pm$     0.02           &      0.27       $\pm$     0.08           &      -0.04       $\pm$     0.11           &      3.1       $\pm$      0.4          \\

\hline
\hline
\end{tabular}
\end{table*}

\begin{table}
\caption[Age80]{Means, gradients, and zero points of the age profiles shown by the black dashed lines (based on  the 20\% percentile of the distribution of the age values in the bins) in Figs. \ref{fig:GradientMN}
 to \ref{fig:GradientDisk}. The different regions are defined in the text and Fig. \ref{fig:regions}. }
\label{tab:Age80}
\begin{tabular}{l c c c}
\hline
\hline
Region    & Mean age       & Age grad.       & Age(r=0)\\
          & [Gyr]          & [Gyr/kpc]       & [Gyr]\\
\hline
\hline
 InBulge  & 10.9 $\pm$ 0.5 &  0.5  $\pm$ 2.0 & 10.9  $\pm$ 0.4 \\
 OutBulge & 10.7 $\pm$ 0.6 &  2.2  $\pm$ 0.6 &  9.4  $\pm$ 0.4 \\
 Bar      &  9.7 $\pm$ 1.0 & -1.2  $\pm$ 0.3 & 10.9  $\pm$ 0.4 \\
 B/P      & 11.1 $\pm$ 0.2 &  0.4  $\pm$ 0.1 & 10.5  $\pm$ 0.2 \\
 Disk     & 3.5  $\pm$ 2.1 &  0.1  $\pm$ 0.8 &  3.0  $\pm$ 3.2 \\
\hline
\hline
\end{tabular}
\end{table}

\begin{table}
 \caption[Model parameters]{Parameters of the composite stellar populations models.}
\label{tab:ParamDB}
\begin{tabular}{l l}
\hline
\hline
Parameter   & Value \\
\hline
\hline
$Age_{CB}$       & 13 Gyr \\
$Age_{CB}$ 20\%  & 11 Gyr \\
$[Z/H]_{CB}(0)$  & 0.35 dex\\
$a_{CB}$         & -0.11 dex/dex\\
$[\alpha/Fe]_{CB}$ & 0.35 dex\\
\hline
$Age_{Bar}$       & 12.2 Gyr\\
$Age_{Bar}$ 20\%  & 9 Gyr\\
$[Z/H]_{Bar}$ & 0.01 dex\\
$[\alpha/Fe]_{Bar}(0)$ & 0.35 dex\\
$a_{Bar}$     & -0.05 dex/dex\\
\hline
$Age_{B/P}$       & 13 Gyr\\
$Age_{B/P}$ 20\%  & 11 Gyr\\
$[Z/H]_{B/P}$ & -0.09 dex\\
$[\alpha/Fe]_{B/P}(0)$ & 0.35 dex\\
$a_{B/P}$     & -0.05 dex/dex\\
\hline
$Age_{Disk}$  &  7 Gyr\\
$Age_{Disk}$ 20\% & 5 Gyr\\
$[Z/H]_{B/P}$ &  0.01 dex\\
$[\alpha/Fe]_{B/P}(0)$ & 0.35 dex\\
$a_{Disk}$     & -0.05 dex/dex\\
\hline
\end{tabular}
\end{table}

It is worth comparing the properties of the M31 B/P bulge we
  discussed above with those of the MW bulge. To start,
  the MW has a strong B/P shape \citep{Wegg13}, while this is not the
  case for M31 \citep{Blana18}. The vertical metallicity gradient in
  the MW is $\approx -0.4$ dex/kpc \citep{Gonzalez2011} and is
  believed to be due to the fact that stars with higher metallicities
  are concentrated more to the Galactic plane and contribute less at
  high latitudes \citep{Zoccali08,Ness13,Portail17b}. In M31, we
  measure a much weaker metallicity gradient (-0.1 dex/kpc) orthogonal
  to the bar 100 arcsec away from the center. Both the MW and M31
  B/P structures have stellar populations that are $\alpha-$enhanced
  \citep{Ness13}. Finally, we derive mean old ages for the stellar
  populations of the bulge region of M31 and acknowledge the presence
  of a $\approx 20$\% of stars younger than 10 Gyr.  For the MW bulge,
  color-magnitude diagrams suggest that the bulge stars of the MW are
  mostly very old, $\approx 10$ Gyr \citep{Clarkson11}, and
  spectroscopy of microlensed dwarf stars points to a fraction of at
  least 18\% younger than 5 Gyr \citep{Bensby17}.

\subsection{Possible formation scenario}

The stellar population analysis presented in the previous section
suggests that the inner 600 arcsec of M31 are made of three
components: an old (11-13 Gyr), metal-rich ([Z/H]$\approx 0.35$ dex in
the center with a negative gradient) and $\alpha-$element overabundant
CB; an almost coeval B/P bulge and a slightly younger bar. The B/P
bulge and the bar have similar $\alpha-$element overabundance
declining slightly outward and subsolar metallicity, which is enhanced
to solar values in the bar region. At larger distances from the center
the stars of the disk of M31 become dominant. They are younger, have
solar metallicity, and are $\alpha-$element overabundant (but see the
caveat at the end of this section).

Given the presence of dust and projection effects, firm conclusions
can be reached only through the projection and analysis of a
three-dimensional model, which we plan for a future paper, following
the approach of \citet{Blana18}.  However, assuming that what is
presented above holds, a possible formation scenario could be the
following. An early, monolithic and quick appearance of the CB, could
explain the metallicity gradient and the lack of iron implied by the
absence of enrichment from supernovae of type Ia observed in that
component. During this phase the formation of a primeval disk is also
going on, which is transformed on a dynamical timescale of the order
of 1 Gyr by the bar formation and buckling into a B/P bulge
\citep{Blana16a}. At this stage the gas is funnelled toward the bar
region, where star formation continues, reducing the mean age of the
stellar population by $\approx 1$ Gyr. Also, the chemical
self-enrichment of the gas is enhanced in the bar region, where the
deeper gravitational potential retains the gas more efficiently,
causing an increase of the metallicity observed along the bar. Gas
must have been available to sustain further star formation and explain
the presence of the $\approx 20$\% of stars younger than 8 Gyr found
by \citet{Dong2018}. This is particularly true in the disk region
probed by our observations, where younger mean ages are
observed. There is an inconsistency in this picture, however. The
radial gradient in [$\alpha$/Fe] that is required by our model along
the bar and in the B/P bulge should imply a more extended period of
star formation as the distance from the center increases. This
contrasts with our assumption of a constant radial age for both the
bar and B/P regions, which we need to follow the observed age profiles
with our two-component model. But our age profiles are probably not
accurate enough to detect this effect, given the spatial spread of
measured ages discussed above. Also, Fig. \ref{fig:correlation} points
to the presence of spectra tagged with ages younger than 6 Gyr and
high [$\alpha$/Fe] overabundance. But the stars in the bulge of the MW
that are younger than 6 Gyr are less $\alpha-$element enhanced (0.126
dex) than those older than 6 Gyr \citep[0.215 dex;][]{Schultheis2017}.
However, the procedure described in Sect.  \ref{spanalysis}, where we
mix SSPs according to the mass distributions of \citet{Dong2018},
shows that we are bound to measure [$\alpha$/Fe] overabundances
systematically biased high owing to the presence of the old and
$\alpha-$element overabundant component that dominates in mass the
younger, possibly less $\alpha-$element overabundant stellar
population.

\section{Conclusions}
\label{sec:Conclusions}

We measured the Lick absorption indices \citep{Trager98} using the M31
spectra presented in Paper I and fit the stellar populations by
applying the method of \citetalias{Saglia10}. We distinguished between
the inner 600 arcsec, where the photometric bulge dominates, and the
outer region of M31, where the disk is prominent. We mapped the
regions where the dust absorption is less of a problem and used these
to trace the average stellar population properties of the classical
and B/P bulge, the bar, and the disk of M31. The main results
from this analysis are

\begin{enumerate}[(i)]
\item H$\beta$ is approximately constant over the whole photometric
  bulge region, showing gradients due to the dust distribution and
  increasing in the disk.  Mgb and several iron indices are enhanced
  along the bar. The presence of dust affects the overall topology of
  the line maps, by masking the light coming from the inner regions of
  M31, north of the disk major axis.
\item The stellar populations are fairly old (10-12 Gyr); 80\% 
of our measurements have ages larger than 10 Gyr. This mirrors
  the result of \citet{Dong2018}, who conclude that 80\% of the stars
  outside 100 arcsec from the center are older than 8 Gyr analyzing
  the bright end of color-magnitude diagrams.
The stars of the disk are younger by 5-6 Gyr.
\item Along the bar the metallicity is solar with a peak of 0.35 dex
  in the very center.  A negative metallicity gradient is
    present, tracing the stellar population of the CB,
  in broad agreement with \citet{Dong2018}.  The B/P bulge has
  subsolar metallicity.
\item The $\alpha/Fe$ overabundance is around 0.25 dex everywhere,
  peaking to 0.35 dex in the CB region.
\item The $M/L_V$ value is $\approx 4.4-4.7 M_\odot/L_\odot$ for the old
  populations of the bar and the classical and B/P bulge, and
  smaller ($\le 3 M_\odot/L_\odot$) in the disk. This supports the
  assumption by \citet{Blana18} of a constant
  mass-to-light ratio for the bar, classical, and B/P bulge
  components.
\end{enumerate} 

We investigated whether the presence of a (young) rotating disk
component in the inner 600 arcsec of M31 with a light distribution
following the photometric major axis decomposition of
\citet{Kormendy99} is compatible with the kinematics and stellar
populations we measured. While kinematically possible, a disk as young
as 4 Gyr is not present in the inner 600 arcsec of M31.
 
We propose a two-phase formation scenario for the inner region of
  M31. A classical stellar bulge forms together with a primordial
  disk from a quasi-monolithic collapse or violent instability of the
  turbulent inner proto-disk. On larger scales the proto-disk develops
  a bar. The bar buckles and transforms the proto-disk into the
B/P bulge. The stars in this region are mainly old.
  However, star formation went on further not only in the disk region,
  but also in the inner 600 arcsec of M31, more intensively in the bar.

In future papers we will explore how a realistic three-dimensional
model taking into account the dust distribution of M31 \citep[as in][]{Blana18}
can reproduce quantitatively our findings. This
might shed further light on the sequence of events that built the
  central regions of M31.  Moreover, we will estimate the rate of
microlensing events expected from the self-lensing of the different
populations of the central region of M31.

\bigskip
{\bf Acknowledgements:} We thank the anonymous referee for a
constructive critical reading of the manuscript.

\bibliographystyle{aa}
\bibliography{Doktorarbeit.bib}

\begin{thebibliography}{79}
\expandafter\ifx\csname natexlab\endcsname\relax\def\natexlab#1{#1}\fi

\bibitem[{{Athanassoula}(2013)}]{Athanassoula13}
{Athanassoula}, E. 2013, {Bars and secular evolution in disk galaxies:
  Theoretical input}, ed. J.~{Falc{\'o}n-Barroso} \& J.~H. {Knapen}, 305

\bibitem[{{Athanassoula} \& {Beaton}(2006)}]{Athanassoula06}
{Athanassoula}, E. \& {Beaton}, R.~L. 2006, \mnras, 370, 1499

\bibitem[{{Athanassoula} {et~al.}(2017){Athanassoula}, {Rodionov}, \&
  {Prantzos}}]{Athanassoula17}
{Athanassoula}, E., {Rodionov}, S.~A., \& {Prantzos}, N. 2017, \mnras, 467, L46

\bibitem[{{Beaton} {et~al.}(2007){Beaton}, {Majewski}, {Guhathakurta},
  {Skrutskie}, {Cutri}, {Good}, {Patterson}, {Athanassoula}, \&
  {Bureau}}]{Beaton07}
{Beaton}, R.~L., {Majewski}, S.~R., {Guhathakurta}, P., {et~al.} 2007, \apjl,
  658, L91

\bibitem[{{Bensby} {et~al.}(2017){Bensby}, {Feltzing}, {Gould}, {Yee},
  {Johnson}, {Asplund}, {Mel{\'e}ndez}, {Lucatello}, {Howes}, {McWilliam},
  {Udalski}, {Szyma{\'n}ski}, {Soszy{\'n}ski}, {Poleski}, {Wyrzykowski},
  {Ulaczyk}, {Koz{\l}owski}, {Pietrukowicz}, {Skowron}, {Mr{\'o}z}, {Pawlak},
  {Abe}, {Asakura}, {Bhattacharya}, {Bond}, {Bennett}, {Hirao}, {Nagakane},
  {Koshimoto}, {Sumi}, {Suzuki}, \& {Tristram}}]{Bensby17}
{Bensby}, T., {Feltzing}, S., {Gould}, A., {et~al.} 2017, \aap, 605, A89

\bibitem[{{Bla\~na} {et~al.}({2018}){Bla\~na}, {Gerhard}, {Wegg}, {Portail},
  {Opitsch}, {Saglia}, {Fabricius}, {Erwin}, \& {Bender}}]{Blana18}
{Bla\~na}, M., {Gerhard}, O., {Wegg}, C., {et~al.} {2018}, submitted to \mnras

\bibitem[{{Bla{\~n}a D{\'{\i}}az} {et~al.}(2017){Bla{\~n}a D{\'{\i}}az},
  {Wegg}, {Gerhard}, {Erwin}, {Portail}, {Opitsch}, {Saglia}, \&
  {Bender}}]{Blana16a}
{Bla{\~n}a D{\'{\i}}az}, M., {Wegg}, C., {Gerhard}, O., {et~al.} 2017, \mnras,
  466, 4279

\bibitem[{{Bruzual} {et~al.}(2017){Bruzual}, {Gladis Magris}, \&
  {Hern{\'a}ndez-P{\'e}rez}}]{Bruzual2017}
{Bruzual}, G., {Gladis Magris}, C., \& {Hern{\'a}ndez-P{\'e}rez}, F. 2017, in
  IAU Symposium, Vol. 321, Formation and Evolution of Galaxy Outskirts, ed.
  A.~{Gil de Paz}, J.~H. {Knapen}, \& J.~C. {Lee}, 96--98

\bibitem[{{Bureau} \& {Athanassoula}(1999)}]{Bureau99}
{Bureau}, M. \& {Athanassoula}, E. 1999, \apj, 522, 686

\bibitem[{{Burstein} {et~al.}(1984){Burstein}, {Faber}, {Gaskell}, \&
  {Krumm}}]{Burstein84}
{Burstein}, D., {Faber}, S.~M., {Gaskell}, C.~M., \& {Krumm}, N. 1984, \apj,
  287, 586

\bibitem[{{Burstein} {et~al.}(1986){Burstein}, {Faber}, \&
  {Gonzalez}}]{Burstein86}
{Burstein}, D., {Faber}, S.~M., \& {Gonzalez}, J.~J. 1986, \aj, 91, 1130

\bibitem[{{Cacho} {et~al.}(2014){Cacho}, {S{\'a}nchez-Bl{\'a}zquez}, {Gorgas},
  \& {P{\'e}rez}}]{Cacho14}
{Cacho}, R., {S{\'a}nchez-Bl{\'a}zquez}, P., {Gorgas}, J., \& {P{\'e}rez}, I.
  2014, \mnras, 442, 2496

\bibitem[{{Cappellari} \& {Copin}(2003)}]{Cappellari03}
{Cappellari}, M. \& {Copin}, Y. 2003, \mnras, 342, 345

\bibitem[{{Cheung} {et~al.}(2013){Cheung}, {Athanassoula}, {Masters}, {Nichol},
  {Bosma}, {Bell}, {Faber}, {Koo}, {Lintott}, {Melvin}, {Schawinski}, {Skibba},
  \& {Willett}}]{Cheung13}
{Cheung}, E., {Athanassoula}, E., {Masters}, K.~L., {et~al.} 2013, \apj, 779,
  162

\bibitem[{{Cheung} {et~al.}(2015){Cheung}, {Trump}, {Athanassoula}, {Bamford},
  {Bell}, {Bosma}, {Cardamone}, {Casteels}, {Faber}, {Fang}, {Fortson},
  {Kocevski}, {Koo}, {Laine}, {Lintott}, {Masters}, {Melvin}, {Nichol},
  {Schawinski}, {Simmons}, {Smethurst}, \& {Willett}}]{Cheung15}
{Cheung}, E., {Trump}, J.~R., {Athanassoula}, E., {et~al.} 2015, \mnras, 447,
  506

\bibitem[{{Chung} \& {Bureau}(2004)}]{Chung04}
{Chung}, A. \& {Bureau}, M. 2004, \aj, 127, 3192

\bibitem[{{Clarkson} {et~al.}(2011){Clarkson}, {Sahu}, {Anderson}, {Rich},
  {Smith}, {Brown}, {Bond}, {Livio}, {Minniti}, {Renzini}, \&
  {Zoccali}}]{Clarkson11}
{Clarkson}, W.~I., {Sahu}, K.~C., {Anderson}, J., {et~al.} 2011, \apj, 735, 37

\bibitem[{{Coelho} \& {Gadotti}(2011)}]{Coelho11}
{Coelho}, P. \& {Gadotti}, D.~A. 2011, \apjl, 743, L13

\bibitem[{{Dalcanton} {et~al.}(2015){Dalcanton}, {Fouesneau}, {Hogg}, {Lang},
  {Leroy}, {Gordon}, {Sandstrom}, {Weisz}, {Williams}, {Bell}, {Dong},
  {Gilbert}, {Gouliermis}, {Guhathakurta}, {Lauer}, {Schruba}, {Seth}, \&
  {Skillman}}]{Dalcanton15}
{Dalcanton}, J.~J., {Fouesneau}, M., {Hogg}, D.~W., {et~al.} 2015, \apj, 814, 3

\bibitem[{{de Lorenzi} {et~al.}(2007){de Lorenzi}, {Debattista}, {Gerhard}, \&
  {Sambhus}}]{deLorenzi07}
{de Lorenzi}, F., {Debattista}, V.~P., {Gerhard}, O., \& {Sambhus}, N. 2007,
  \mnras, 376, 71

\bibitem[{{Debattista} {et~al.}(2017){Debattista}, {Ness}, {Gonzalez},
  {Freeman}, {Zoccali}, \& {Minniti}}]{Debattista17}
{Debattista}, V.~P., {Ness}, M., {Gonzalez}, O.~A., {et~al.} 2017, \mnras, 469,
  1587

\bibitem[{{Di Matteo}(2016)}]{DiMatteo2016}
{Di Matteo}, P. 2016, \pasa, 33, e027

\bibitem[{{Di Matteo} {et~al.}(2015){Di Matteo}, {G{\'o}mez}, {Haywood},
  {Combes}, {Lehnert}, {Ness}, {Snaith}, {Katz}, \& {Semelin}}]{DiMatteo2015}
{Di Matteo}, P., {G{\'o}mez}, A., {Haywood}, M., {et~al.} 2015, \aap, 577, A1

\bibitem[{{Di Matteo} {et~al.}(2013){Di Matteo}, {Haywood}, {Combes},
  {Semelin}, \& {Snaith}}]{DiMatteo13}
{Di Matteo}, P., {Haywood}, M., {Combes}, F., {Semelin}, B., \& {Snaith}, O.~N.
  2013, \aap, 553, A102

\bibitem[{{Dong} {et~al.}(2018){Dong}, {Olsen}, {Lauer}, {Saha}, {Li},
  {Garc{\'{\i}}a-Benito}, \& {Sch{\"o}del}}]{Dong2018}
{Dong}, H., {Olsen}, K., {Lauer}, T., {et~al.} 2018, \mnras, 478, 5379

\bibitem[{{Draine} {et~al.}(2014){Draine}, {Aniano}, {Krause}, {Groves},
  {Sandstrom}, {Braun}, {Leroy}, {Klaas}, {Linz}, {Rix}, {Schinnerer},
  {Schmiedeke}, \& {Walter}}]{Draine2014}
{Draine}, B.~T., {Aniano}, G., {Krause}, O., {et~al.} 2014, \apj, 780, 172

\bibitem[{{Draine} \& {Li}(2007)}]{Draine2007}
{Draine}, B.~T. \& {Li}, A. 2007, \apj, 657, 810

\bibitem[{{Erwin} {et~al.}(2015){Erwin}, {Saglia}, {Fabricius}, {Thomas},
  {Nowak}, {Rusli}, {Bender}, {Vega Beltr{\'a}n}, \& {Beckman}}]{Erwin15}
{Erwin}, P., {Saglia}, R.~P., {Fabricius}, M., {et~al.} 2015, \mnras, 446, 4039

\bibitem[{{Faber} {et~al.}(1985){Faber}, {Friel}, {Burstein}, \&
  {Gaskell}}]{Faber85}
{Faber}, S.~M., {Friel}, E.~D., {Burstein}, D., \& {Gaskell}, C.~M. 1985,
  \apjs, 57, 711

\bibitem[{{Fathi} \& {Peletier}(2003)}]{Fathi03}
{Fathi}, K. \& {Peletier}, R.~F. 2003, \aap, 407, 61

\bibitem[{{Fragkoudi} {et~al.}(2017){Fragkoudi}, {Di Matteo}, {Haywood},
  {Khoperskov}, {Gomez}, {Schultheis}, {Combes}, \& {Semelin}}]{Fragkoudi17}
{Fragkoudi}, F., {Di Matteo}, P., {Haywood}, M., {et~al.} 2017, \aap, 607, L4

\bibitem[{{Friedli} {et~al.}(1994){Friedli}, {Benz}, \&
  {Kennicutt}}]{Friedli94}
{Friedli}, D., {Benz}, W., \& {Kennicutt}, R. 1994, \apjl, 430, L105

\bibitem[{{Gadotti} \& {dos Anjos}(2001)}]{Gadotti01}
{Gadotti}, D.~A. \& {dos Anjos}, S. 2001, \aj, 122, 1298

\bibitem[{{Gonzalez} {et~al.}(2017){Gonzalez}, {Debattista}, {Ness}, {Erwin},
  \& {Gadotti}}]{Gonzalez17}
{Gonzalez}, O.~A., {Debattista}, V.~P., {Ness}, M., {Erwin}, P., \& {Gadotti},
  D.~A. 2017, \mnras, 466, L93

\bibitem[{{Gonzalez} {et~al.}(2013){Gonzalez}, {Rejkuba}, {Zoccali}, {Valent},
  {Minniti}, \& {Tobar}}]{Gonzalez13}
{Gonzalez}, O.~A., {Rejkuba}, M., {Zoccali}, M., {et~al.} 2013, \aap, 552, A110

\bibitem[{{Gonzalez} {et~al.}(2011){Gonzalez}, {Rejkuba}, {Zoccali}, {Valenti},
  \& {Minniti}}]{Gonzalez2011}
{Gonzalez}, O.~A., {Rejkuba}, M., {Zoccali}, M., {Valenti}, E., \& {Minniti},
  D. 2011, \aap, 534, A3

\bibitem[{{Gregersen} {et~al.}(2015){Gregersen}, {Seth}, {Williams}, {Lang},
  {Dalcanton}, {Girardi}, {Skillman}, {Bell}, {Dolphin}, {Fouesneau},
  {Guhathakurta}, {Hamren}, {Johnson}, {Kalirai}, {Lewis}, {Monachesi}, \&
  {Olsen}}]{Gregersen15}
{Gregersen}, D., {Seth}, A.~C., {Williams}, B.~F., {et~al.} 2015, \aj, 150, 189

\bibitem[{{Knapen} {et~al.}(1995){Knapen}, {Beckman}, {Heller}, {Shlosman}, \&
  {de Jong}}]{Knapen95}
{Knapen}, J.~H., {Beckman}, J.~E., {Heller}, C.~H., {Shlosman}, I., \& {de
  Jong}, R.~S. 1995, \apj, 454, 623

\bibitem[{{Kormendy}(2013)}]{Kormendy13}
{Kormendy}, J. 2013, {Secular Evolution in Disk Galaxies}, ed.
  J.~{Falc{\'o}n-Barroso} \& J.~H. {Knapen}, 1

\bibitem[{{Kormendy} \& {Bender}(1999)}]{Kormendy99}
{Kormendy}, J. \& {Bender}, R. 1999, \apj, 522, 772

\bibitem[{{Kormendy} \& {Kennicutt}(2004)}]{Kormendy04}
{Kormendy}, J. \& {Kennicutt}, Jr., R.~C. 2004, \araa, 42, 603

\bibitem[{{Kroupa}(2001)}]{Kroupa01}
{Kroupa}, P. 2001, \mnras, 322, 231

\bibitem[{{Kubryk} {et~al.}(2013){Kubryk}, {Prantzos}, \&
  {Athanassoula}}]{Kubryk13}
{Kubryk}, M., {Prantzos}, N., \& {Athanassoula}, E. 2013, \mnras, 436, 1479

\bibitem[{{Lee} {et~al.}(2012){Lee}, {Riffeser}, {Koppenhoefer}, {Seitz},
  {Bender}, {Hopp}, {G{\"o}ssl}, {Saglia}, {Snigula}, {Sweeney}, {Burgett},
  {Chambers}, {Grav}, {Heasley}, {Hodapp}, {Kaiser}, {Magnier}, {Morgan},
  {Price}, {Stubbs}, {Tonry}, \& {Wainscoat}}]{Lee12}
{Lee}, C.-H., {Riffeser}, A., {Koppenhoefer}, J., {et~al.} 2012, \aj, 143, 89

\bibitem[{{Lindblad}(1956)}]{Lindblad56}
{Lindblad}, B. 1956, Stockholms Observatoriums Annaler, 19

\bibitem[{{Maraston}(1998)}]{Maraston98}
{Maraston}, C. 1998, \mnras, 300, 872

\bibitem[{{Maraston}(2005)}]{Maraston05}
{Maraston}, C. 2005, \mnras, 362, 799

\bibitem[{{Martinez-Valpuesta} \& {Gerhard}(2013)}]{Valpuesta13}
{Martinez-Valpuesta}, I. \& {Gerhard}, O. 2013, \apjl, 766, L3

\bibitem[{{Minchev} \& {Famaey}(2010)}]{Minchev10}
{Minchev}, I. \& {Famaey}, B. 2010, \apj, 722, 112

\bibitem[{{Moorthy} \& {Holtzman}(2006)}]{Moorthy06}
{Moorthy}, B.~K. \& {Holtzman}, J.~A. 2006, \mnras, 371, 583

\bibitem[{{Ness} {et~al.}(2013){Ness}, {Freeman}, {Athanassoula},
  {Wylie-de-Boer}, {Bland-Hawthorn}, {Asplund}, {Lewis}, {Yong}, {Lane}, \&
  {Kiss}}]{Ness13}
{Ness}, M., {Freeman}, K., {Athanassoula}, E., {et~al.} 2013, \mnras, 430, 836

\bibitem[{{Opitsch}(2016)}]{Opitsch2016}
{Opitsch}, M. 2016, PhD thesis, Ludwig-Maximilians-Universit\"at M\"unchen,
  Germany

\bibitem[{{Opitsch} {et~al.}(2018){Opitsch}, {Fabricius}, {Saglia}, {Bender},
  {Bla{\~n}a}, \& {Gerhard}}]{Opitsch18}
{Opitsch}, M., {Fabricius}, M.~H., {Saglia}, R.~P., {et~al.} 2018, \aap, 611,
  A38

\bibitem[{{P{\'e}rez} \& {S{\'a}nchez-Bl{\'a}zquez}(2011)}]{Perez11}
{P{\'e}rez}, I. \& {S{\'a}nchez-Bl{\'a}zquez}, P. 2011, \aap, 529, A64

\bibitem[{{P{\'e}rez} {et~al.}(2007){P{\'e}rez}, {S{\'a}nchez-Bl{\'a}zquez}, \&
  {Zurita}}]{Perez07}
{P{\'e}rez}, I., {S{\'a}nchez-Bl{\'a}zquez}, P., \& {Zurita}, A. 2007, \aap,
  465, L9

\bibitem[{{P{\'e}rez} {et~al.}(2009){P{\'e}rez}, {S{\'a}nchez-Bl{\'a}zquez}, \&
  {Zurita}}]{Perez09}
{P{\'e}rez}, I., {S{\'a}nchez-Bl{\'a}zquez}, P., \& {Zurita}, A. 2009, \aap,
  495, 775

\bibitem[{{Portail} {et~al.}(2017{\natexlab{a}}){Portail}, {Gerhard}, {Wegg},
  \& {Ness}}]{Portail17}
{Portail}, M., {Gerhard}, O., {Wegg}, C., \& {Ness}, M. 2017{\natexlab{a}},
  \mnras, 465, 1621

\bibitem[{{Portail} {et~al.}(2017{\natexlab{b}}){Portail}, {Wegg}, {Gerhard},
  \& {Ness}}]{Portail17b}
{Portail}, M., {Wegg}, C., {Gerhard}, O., \& {Ness}, M. 2017{\natexlab{b}},
  \mnras, 470, 1233

\bibitem[{{Prugniel} {et~al.}(2007){Prugniel}, {Soubiran}, {Koleva}, \& {Le
  Borgne}}]{Prugniel07}
{Prugniel}, P., {Soubiran}, C., {Koleva}, M., \& {Le Borgne}, D. 2007, VizieR
  Online Data Catalog, 3251, 0

\bibitem[{{Riffeser} {et~al.}(2008){Riffeser}, {Seitz}, \&
  {Bender}}]{Riffeser08}
{Riffeser}, A., {Seitz}, S., \& {Bender}, R. 2008, \apj, 684, 1093

\bibitem[{{Ro{\v s}kar} {et~al.}(2012){Ro{\v s}kar}, {Debattista}, {Quinn}, \&
  {Wadsley}}]{Roskar12}
{Ro{\v s}kar}, R., {Debattista}, V.~P., {Quinn}, T.~R., \& {Wadsley}, J. 2012,
  \mnras, 426, 2089

\bibitem[{{Saglia} {et~al.}(2010){Saglia}, {Fabricius}, {Bender}, {Montalto},
  {Lee}, {Riffeser}, {Seitz}, {Morganti}, {Gerhard}, \& {Hopp}}]{Saglia10}
{Saglia}, R.~P., {Fabricius}, M., {Bender}, R., {et~al.} 2010, \aap, 509, A61,
  {abbreviated as S10 in the text}

\bibitem[{{Sakamoto} {et~al.}(1999){Sakamoto}, {Okumura}, {Ishizuki}, \&
  {Scoville}}]{Sakamoto99}
{Sakamoto}, K., {Okumura}, S.~K., {Ishizuki}, S., \& {Scoville}, N.~Z. 1999,
  \apj, 525, 691

\bibitem[{{S{\'a}nchez-Bl{\'a}zquez} {et~al.}(2011){S{\'a}nchez-Bl{\'a}zquez},
  {Ocvirk}, {Gibson}, {P{\'e}rez}, \& {Peletier}}]{Sanchez-Blazquez11}
{S{\'a}nchez-Bl{\'a}zquez}, P., {Ocvirk}, P., {Gibson}, B.~K., {P{\'e}rez}, I.,
  \& {Peletier}, R.~F. 2011, \mnras, 415, 709

\bibitem[{{Sarajedini} \& {Jablonka}(2005)}]{Sarajedini05}
{Sarajedini}, A. \& {Jablonka}, P. 2005, \aj, 130, 1627

\bibitem[{{Sarzi} {et~al.}(2006){Sarzi}, {Falc{\'o}n-Barroso}, {Davies},
  {Bacon}, {Bureau}, {Cappellari}, {de Zeeuw}, {Emsellem}, {Fathi},
  {Krajnovi{\'c}}, {Kuntschner}, {McDermid}, \& {Peletier}}]{Sarzi06}
{Sarzi}, M., {Falc{\'o}n-Barroso}, J., {Davies}, R.~L., {et~al.} 2006, \mnras,
  366, 1151

\bibitem[{{Schultheis} {et~al.}(2017){Schultheis}, {Rojas-Arriagada},
  {Garc{\'{\i}}a P{\'e}rez}, {J{\"o}nsson}, {Hayden}, {Nandakumar}, {Cunha},
  {Allende Prieto}, {Holtzman}, {Beers}, {Bizyaev}, {Brinkmann}, {Carrera},
  {Cohen}, {Geisler}, {Hearty}, {Fernandez-Tricado}, {Maraston}, {Minnitti},
  {Nitschelm}, {Roman-Lopes}, {Schneider}, {Tang}, {Villanova}, {Zasowski}, \&
  {Majewski}}]{Schultheis2017}
{Schultheis}, M., {Rojas-Arriagada}, A., {Garc{\'{\i}}a P{\'e}rez}, A.~E.,
  {et~al.} 2017, \aap, 600, A14

\bibitem[{{Seidel} {et~al.}(2016){Seidel}, {Falc{\'o}n-Barroso},
  {Mart{\'{\i}}nez-Valpuesta}, {S{\'a}nchez-Bl{\'a}zquez}, {P{\'e}rez},
  {Peletier}, \& {Vazdekis}}]{Seidel16}
{Seidel}, M.~K., {Falc{\'o}n-Barroso}, J., {Mart{\'{\i}}nez-Valpuesta}, I.,
  {et~al.} 2016, \mnras, 460, 3784

\bibitem[{{Sellwood}(2014)}]{Sellwood14}
{Sellwood}, J.~A. 2014, Reviews of Modern Physics, 86, 1

\bibitem[{{Thomas} \& {Davies}(2006)}]{Thomas06}
{Thomas}, D. \& {Davies}, R.~L. 2006, \mnras, 366, 510

\bibitem[{{Thomas} {et~al.}(2003){Thomas}, {Maraston}, \& {Bender}}]{Thomas03}
{Thomas}, D., {Maraston}, C., \& {Bender}, R. 2003, \mnras, 343, 279

\bibitem[{{Thomas} {et~al.}(2011){Thomas}, {Maraston}, \&
  {Johansson}}]{Thomas11}
{Thomas}, D., {Maraston}, C., \& {Johansson}, J. 2011, \mnras, 412, 2183

\bibitem[{{Trager} {et~al.}(1998){Trager}, {Worthey}, {Faber}, {Burstein}, \&
  {Gonz{\'a}lez}}]{Trager98}
{Trager}, S.~C., {Worthey}, G., {Faber}, S.~M., {Burstein}, D., \&
  {Gonz{\'a}lez}, J.~J. 1998, \apjs, 116, 1

\bibitem[{{Vazdekis} {et~al.}(2015){Vazdekis}, {Coelho}, {Cassisi},
  {Ricciardelli}, {Falc{\'o}n-Barroso}, {S{\'a}nchez-Bl{\'a}zquez}, {La
  Barbera}, {Beasley}, \& {Pietrinferni}}]{Vazdekis2015}
{Vazdekis}, A., {Coelho}, P., {Cassisi}, S., {et~al.} 2015, \mnras, 449, 1177

\bibitem[{{Wegg} \& {Gerhard}(2013)}]{Wegg13}
{Wegg}, C. \& {Gerhard}, O. 2013, \mnras, 435, 1874

\bibitem[{{Williams} {et~al.}(2012){Williams}, {Bureau}, \&
  {Kuntschner}}]{Williams12a}
{Williams}, M.~J., {Bureau}, M., \& {Kuntschner}, H. 2012, \mnras, 427, L99

\bibitem[{{Worthey} {et~al.}(1994){Worthey}, {Faber}, {Gonzalez}, \&
  {Burstein}}]{Worthey94}
{Worthey}, G., {Faber}, S.~M., {Gonzalez}, J.~J., \& {Burstein}, D. 1994,
  \apjs, 94, 687

\bibitem[{{Zieleniewski} {et~al.}(2015){Zieleniewski}, {Houghton}, {Thatte}, \&
  {Davies}}]{Zieleniewski15}
{Zieleniewski}, S., {Houghton}, R.~C.~W., {Thatte}, N., \& {Davies}, R.~L.
  2015, \mnras, 452, 597

\bibitem[{{Zoccali} {et~al.}(2008){Zoccali}, {Hill}, {Lecureur}, {Barbuy},
  {Renzini}, {Minniti}, {G{\'o}mez}, \& {Ortolani}}]{Zoccali08}
{Zoccali}, M., {Hill}, V., {Lecureur}, A., {et~al.} 2008, \aap, 486, 177

\end{thebibliography}

\section{Appendix}
\subsection{Comparison with the Lick index measurements of \citetalias{Saglia10}}
\label{sec:Saglia_indices}

\begin{figure}
\resizebox{\hsize}{!}{\includegraphics{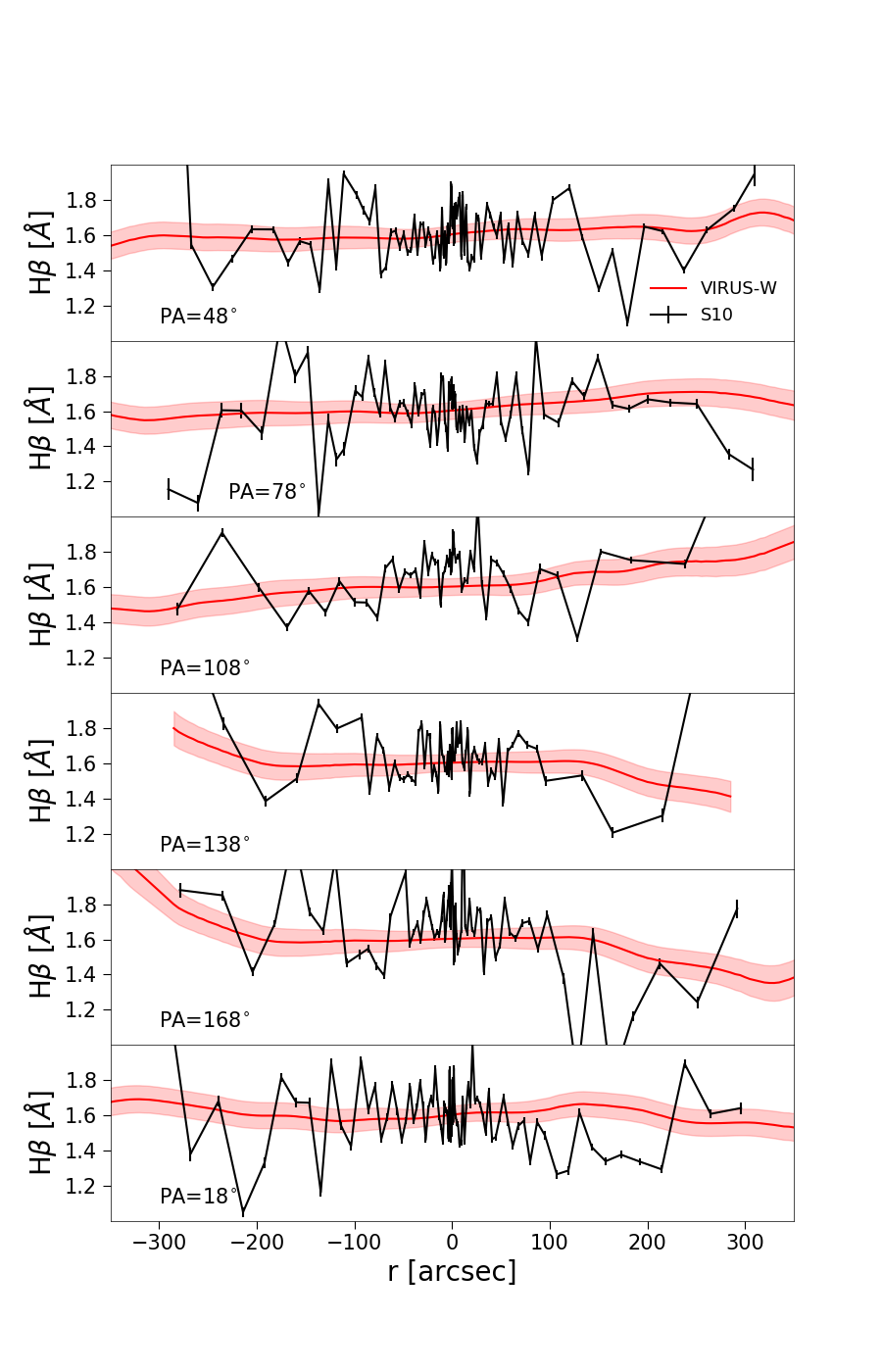}}
\caption[Comparison of H$\beta$ with \citetalias{Saglia10}]{Cuts through the H$\beta$ map from Fig. 
\ref{fig:LickHbeta_map} (red) compared to data from \citetalias{Saglia10} (black).}
\label{fig:Hb_Saglia}
\end{figure}

\begin{figure}
\resizebox{\hsize}{!}{\includegraphics{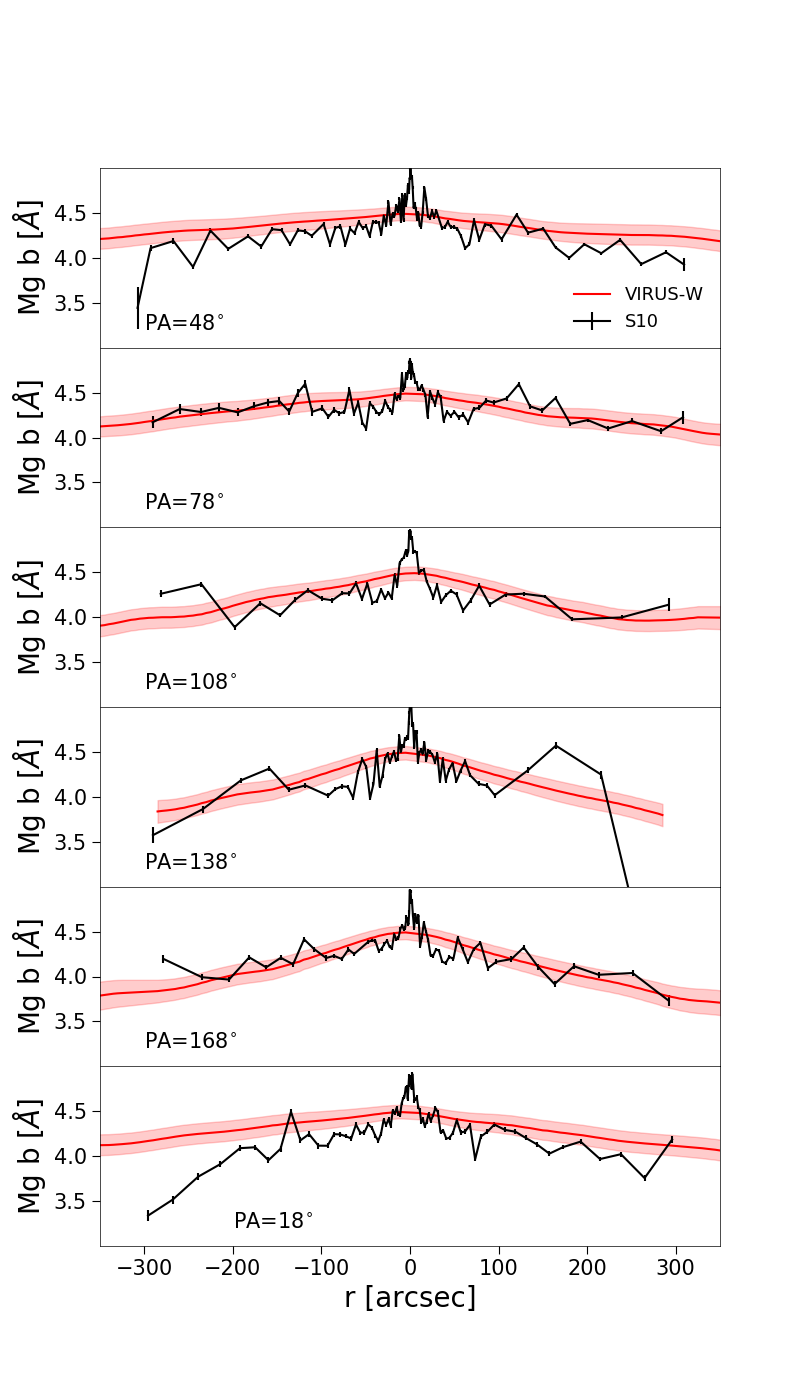}}
\caption[Comparison of Mg b with \citetalias{Saglia10}]{Cuts through the Mg b map from Fig. 
\ref{fig:LickMgb_map} (red) compared to data from \citetalias{Saglia10} (black).}
\label{fig:Mgb_Saglia}
\end{figure}

\begin{figure}
\resizebox{\hsize}{!}{\includegraphics{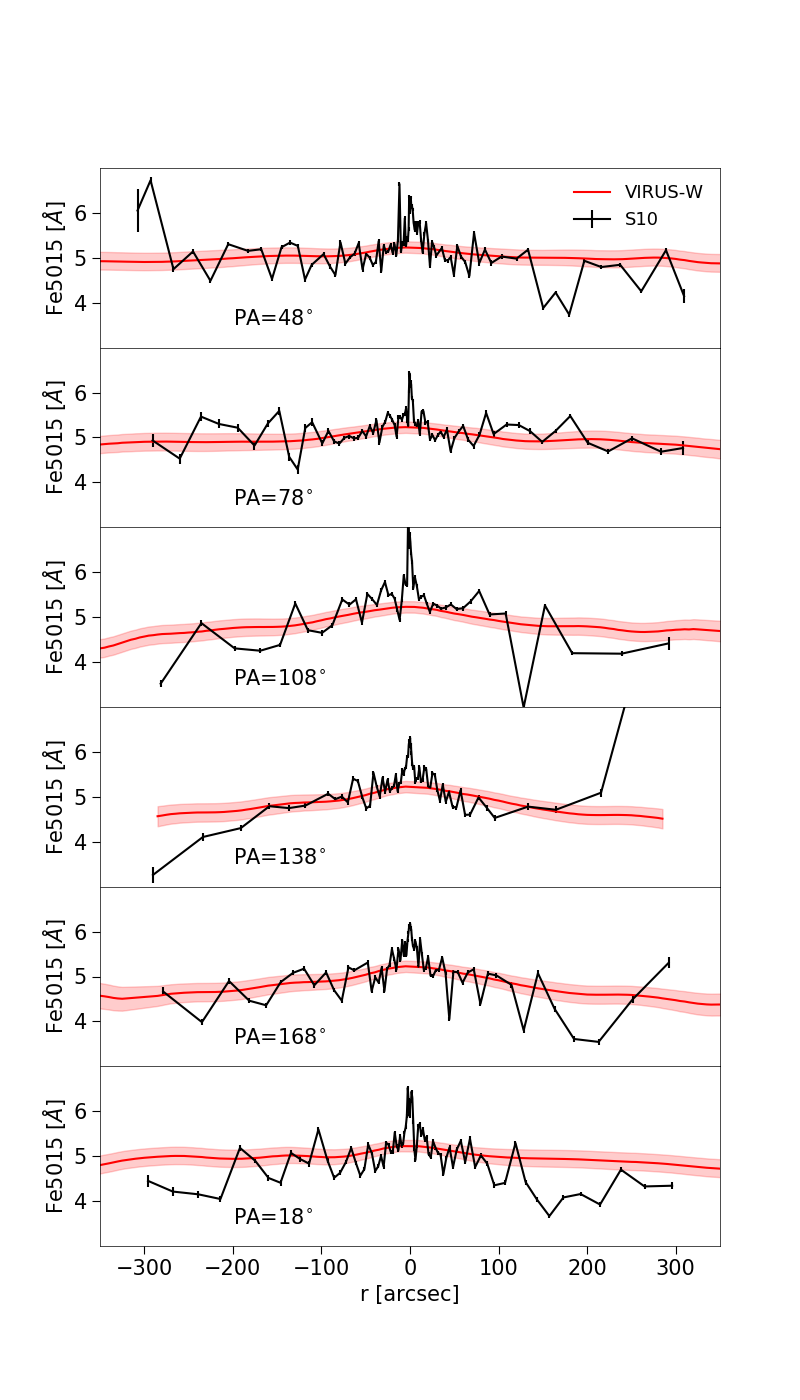}}
\caption[Comparison of Fe5015 with \citetalias{Saglia10}]{Cuts through the Fe5015 map from Fig. 
\ref{fig:LickFe5015_map} (red) compared to data from \citetalias{Saglia10} (black).}
\label{fig:Fe5015_Saglia}
\end{figure}

\begin{figure}
\resizebox{\hsize}{!}{\includegraphics{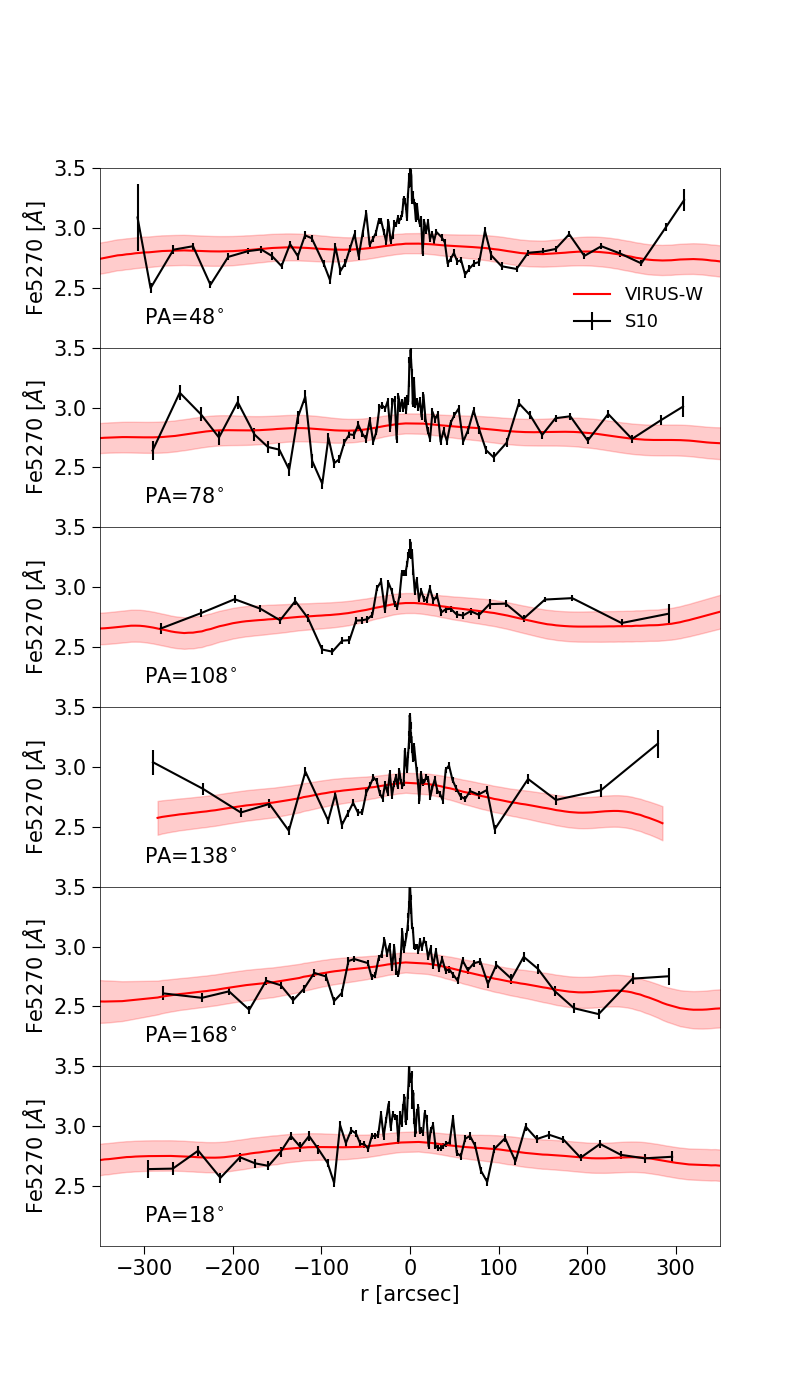}}
\caption[Comparison of Fe5270 with \citetalias{Saglia10}]{Cuts through the Fe5270 map from Fig. 
\ref{fig:LickFe5270_map} (red) compared to data from \citetalias{Saglia10} (black).}
\label{fig:Fe5270_Saglia}
\end{figure}

\begin{figure}
\resizebox{\hsize}{!}{\includegraphics{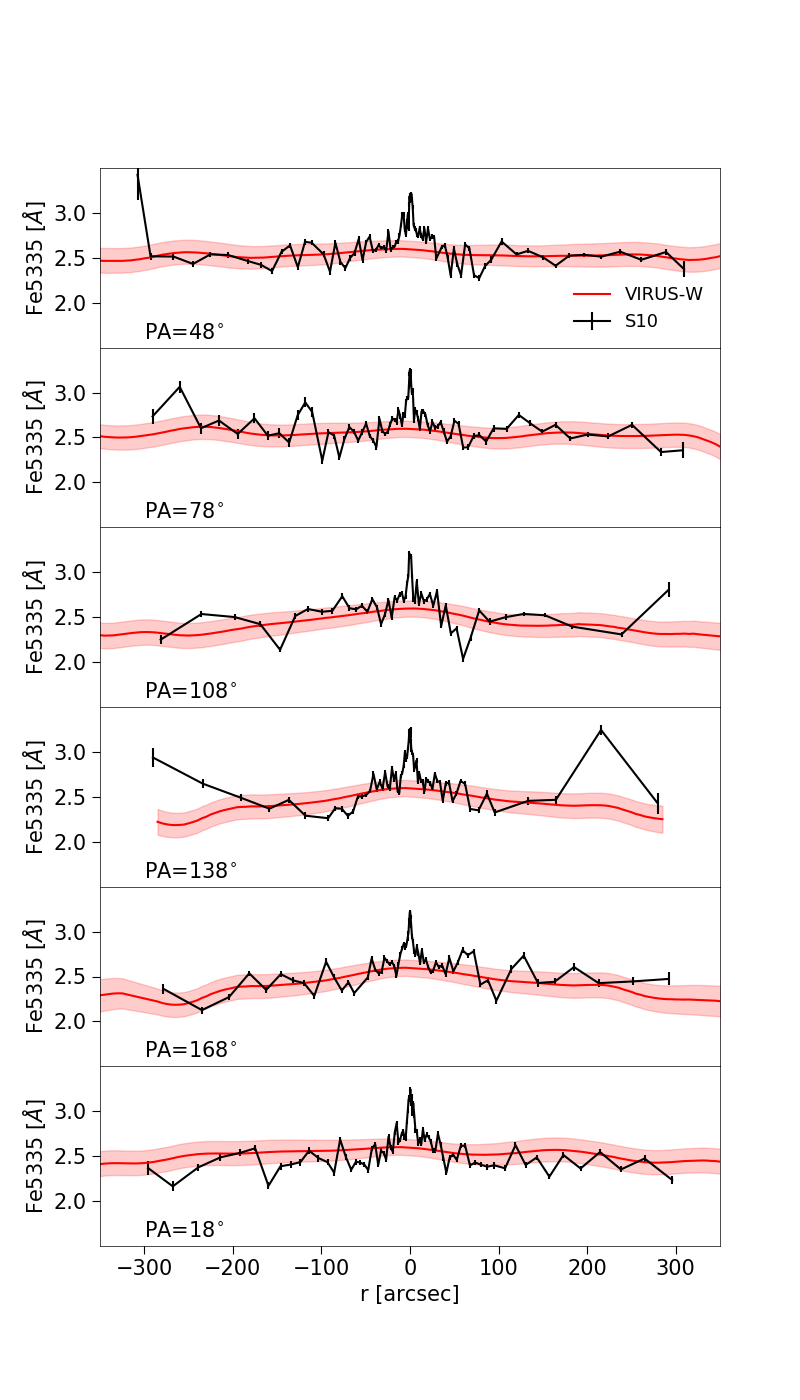}}
\caption[Comparison of Fe5335 with \citetalias{Saglia10}]{Cuts through the Fe5335 map from Fig. 
\ref{fig:LickFe5335_map} (red) compared to data from \citetalias{Saglia10} (black).}
\label{fig:Fe5335_Saglia}
\end{figure}

\begin{figure}
\resizebox{\hsize}{!}{\includegraphics{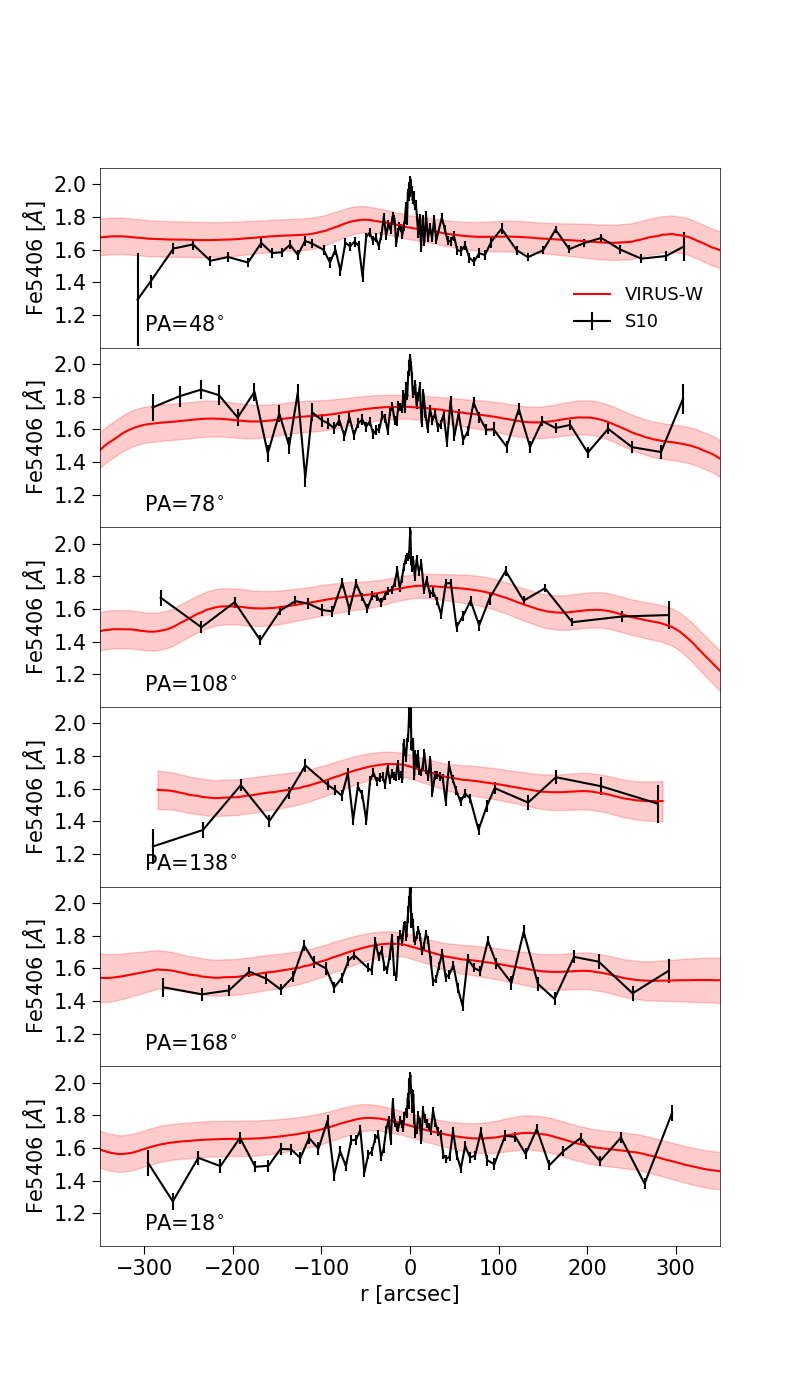}}
\caption[Comparison of Fe5406 with \citetalias{Saglia10}]{Cuts through the Fe5406 map from Fig. 
\ref{fig:LickFe5406_map} (red) compared to data from \citetalias{Saglia10} (black).}
\label{fig:Fe5406_Saglia}
\end{figure}

In Fig. \ref{fig:Hb_Saglia} to \ref{fig:Fe5406_Saglia}, we compare
cuts through our data to the values from \citetalias{Saglia10}. While
the individual measurements sometimes deviate because the two data
sets sample small scale variations of the stellar populations
differently, the general trends are similar; the overall agreement
is within 10$\%$ without systematic shifts. We conclude that our
measurements are on the Lick system.  We do not reproduce the spike in
the very center seen by \citetalias{Saglia10} because the VIRUS-W
data have lower spatial resolution than the \citetalias{Saglia10} data.


\subsection{Comparison with the stellar population measurements of \citetalias{Saglia10}}
\label{sec:Saglia_pop}

\begin{figure}
\resizebox{\hsize}{!}{\includegraphics{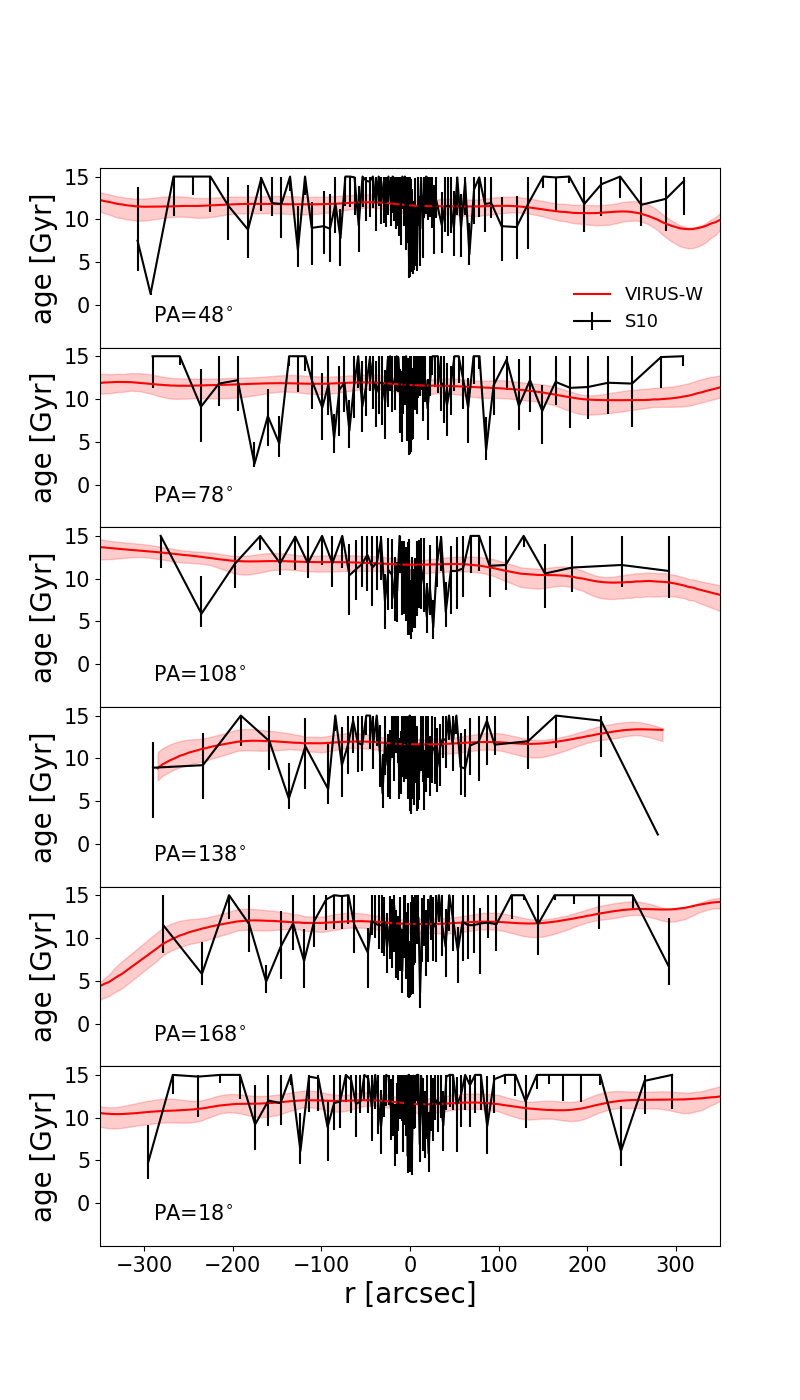}}
\caption[Comparison of age with \citetalias{Saglia10}]{Cuts through our age measurements (red) compared to data from \citetalias{Saglia10} (black).}
\label{fig:age_Saglia}
\end{figure}

\begin{figure}
\resizebox{\hsize}{!}{\includegraphics{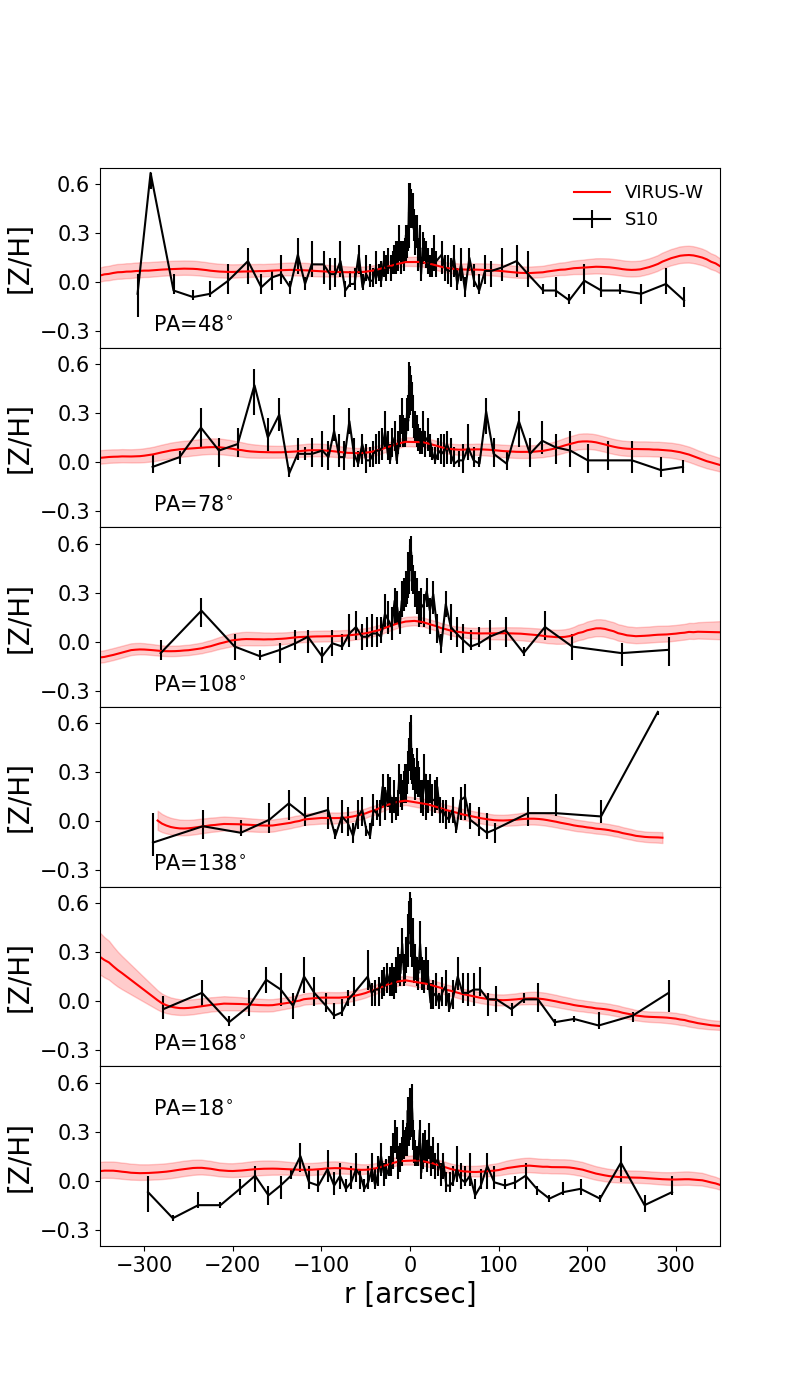}}
\caption[Comparison of metallicity with \citetalias{Saglia10}]{Cuts through our metallicity measurements (red) compared to data from \citetalias{Saglia10} (black).}
\label{fig:ZH_Saglia}
\end{figure}

\begin{figure}
\resizebox{\hsize}{!}{\includegraphics{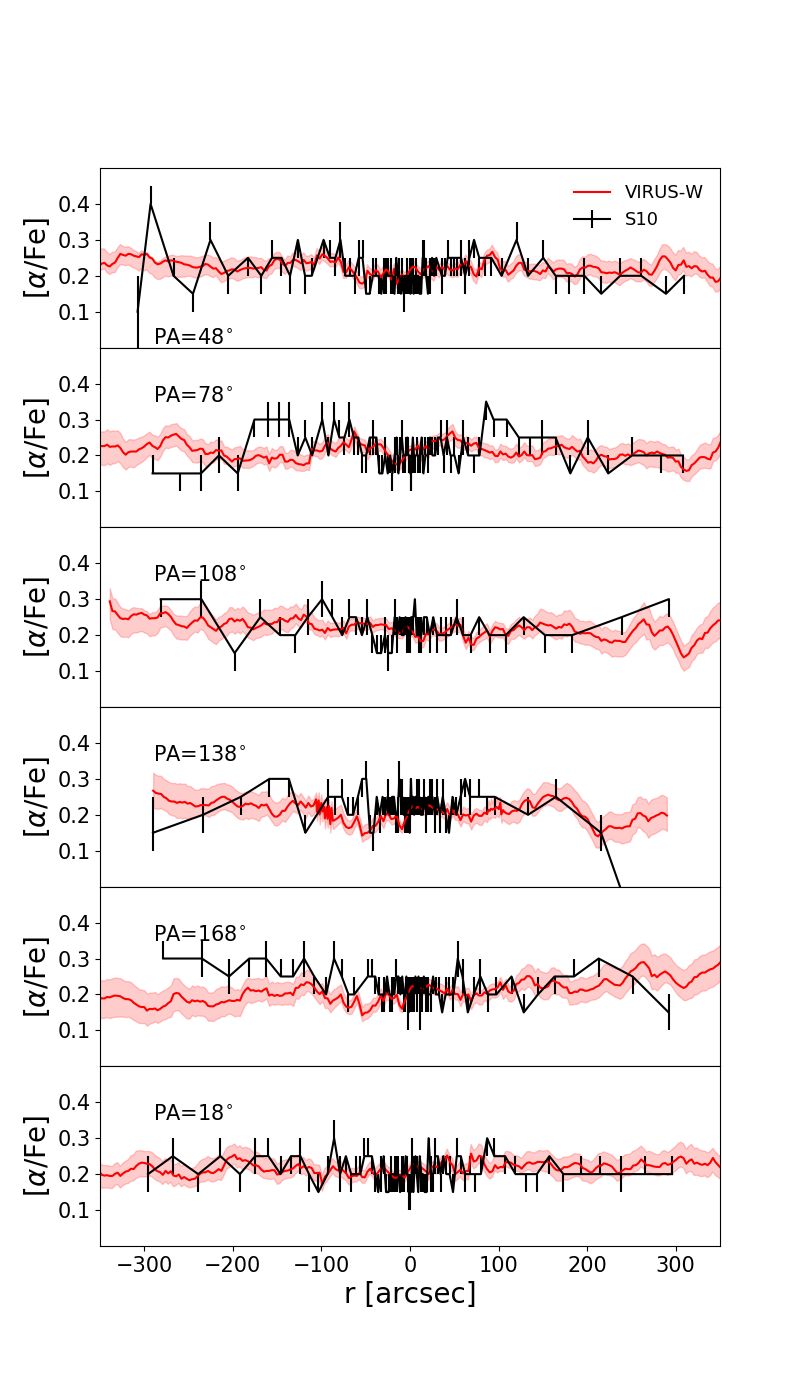}}
\caption[Comparison of $\alpha$/Fe with \citetalias{Saglia10}]{Cuts through our [$\alpha/$Fe] measurements (red) compared to data from \citetalias{Saglia10} (black).}
\label{fig:aFe_Saglia}
\end{figure}

The comparison of our stellar population parameters to those
derived by \citetalias{Saglia10} is presented in Figs.
\ref{fig:age_Saglia} to \ref{fig:aFe_Saglia}. While there is scatter,
in general, both datasets agree within the errors.  Variations of the stellar populations that are too small scale and different spatial
sampling of the present data set and \citetalias{Saglia10} can explain
the differences. The younger population that \citetalias{Saglia10}
find in the innermost 5 arcseconds is not seen in our data. This is, once again,
because the present data set does not have the spatial resolution that
far in.

\subsection{Measured values}

Table \ref{tab:binid} gives the coordinates (relative to the center)
of the spectra corresponding to the bin identification number.  Table
\ref{tab:Lick_indices} presents the format of the tabulated values of
the Lick indices used in this paper (H$\beta$, Mg b, Fe5015, Fe5270,
Fe5355, and Fe5406). Table \ref{tab:Stellar_populations} gives the
resulting stellar population properties (Age, [Z/H], [$\alpha/$Fe], and
$M/L_V$). The complete lists are available online.

\begin{table} 
 \caption{Bin identification.}
\label{tab:binid} 
\begin{tabular}{l l l}
\hline
\hline
Bin number & x[''] & y[''] \\
\hline
\hline
0 &  2.65 &  4.57\\     
1 &  0.08 &  9.14\\     
2 &   2.3 & 11.38\\     
3 &  5.38 &  9.11\\     
4 & -2.92 & 11.16\\     
    ...\\
    \hline
 \end{tabular}
 \end{table}

\begin{table*} 
 \caption{Measured Lick indices}
\label{tab:Lick_indices} 
\begin{tabular}{l l l l l l l}
\hline
\hline
%
%
Bin Number & H$\beta$ [$\AA$] & Mg b [$\AA$] & Fe5015  [$\AA$] & Fe5270  [$\AA$] & Fe5355  [$\AA$] & Fe5406  [$\AA$] \\ 
\hline
\hline
%
0 & 1.64 $\pm$ 0.03 & 4.55 $\pm$ 0.05 & 5.63 $\pm$ 0.09 & 2.98 $\pm$ 0.06 & 2.79 $\pm$ 0.06 & 1.68 $\pm$ 0.05\\
1 & 1.69 $\pm$ 0.04 & 4.45 $\pm$ 0.06 & 5.17 $\pm$ 0.11 & 2.86 $\pm$ 0.07 & 2.67 $\pm$ 0.07 & 1.73 $\pm$ 0.06\\
2 & 1.81 $\pm$ 0.04 & 4.53 $\pm$ 0.06 & 5.49 $\pm$ 0.10 & 2.77 $\pm$ 0.06 & 2.50 $\pm$ 0.07 & 1.51 $\pm$ 0.06\\
3 & 1.48 $\pm$ 0.04 & 4.58 $\pm$ 0.06 & 5.68 $\pm$ 0.10 & 3.08 $\pm$ 0.07 & 2.63 $\pm$ 0.07 & 1.47 $\pm$ 0.06\\
4 & 2.00 $\pm$ 0.03 & 4.52 $\pm$ 0.05 & 5.61 $\pm$ 0.09 & 2.88 $\pm$ 0.06 & 2.55 $\pm$ 0.07 & 1.68 $\pm$ 0.05\\
    ...\\
    \hline
 \end{tabular}
 \end{table*}

\begin{landscape}
\begin{table} 
 \caption{Stellar population parameters.}
\label{tab:Stellar_populations} 
\begin{tabular}{l l l l l l l l l l l l l l l l}
\hline
\hline
%
%
Bin Number & Age  &min Age&max Age&Age20\%&[Z/H]&min [Z/H]& max [Z/H] & [$\alpha/$Fe] & min [$\alpha/$Fe] & max [$\alpha/$Fe] &$M/L_V$& min $M/L_V$ & max $M/L_V$& $M/L_V$20\%\\
           & [Gyr]& [Gyr]&[Gyr]&[Gyr]&  [dex]& [dex]& [dex] & [dex] & [dex] & [dex] &[$M_\odot/L_\odot$]&[$M_\odot/L_\odot$]&[$M_\odot/L_\odot$]&[$M_\odot/L_\odot$] \\
\hline
\hline
%
0 & 11.2 & 10.7 & 11.9 &  10.5 & 0.21 & 0.19 & 0.23 & 0.25 & 0.24 & 0.26 &  4.7 & 4.3  & 4.9 & 4.4\\
1 & 11.6 & 11.0 & 12.6 &  10.6 & 0.13 & 0.09 & 0.15 & 0.25 & 0.24 & 0.26 &  4.6 & 4.2  & 4.8 & 4.1\\
2 &  8.3 &  7.8 &  8.9 &  10.6 & 0.25 & 0.23 & 0.27 & 0.35 & 0.34 & 0.36 &  3.5 & 3.3  & 3.7 & 4.5\\
3 & 14.4 & 13.7 & 15.0 &  10.6 & 0.11 & 0.09 & 0.13 & 0.25 & 0.24 & 0.26 &  5.6 & 5.2  & 5.8 & 4.1\\
4 &  5.0 &  4.8 &  5.4 &  10.6 & 0.39 & 0.37 & 0.40 & 0.35 & 0.34 & 0.36 &  2.4 & 2.3  & 2.6 & 4.9\\
    ...\\
    \hline
 \end{tabular}
 \end{table}
\end{landscape}
\end{document}